\documentclass[%
 reprint,
 amsmath,amssymb,
 aps,
]{revtex4-2}

\usepackage{graphicx}
\usepackage{dcolumn}
\usepackage{bm}
\usepackage{color}
\usepackage{graphicx}
\usepackage{wrapfig}
\usepackage{float}
\DeclareMathOperator{\sgn}{sgn}
\usepackage{hyperref}

\begin{document}

\preprint{APS/123-QED}
\title{A combinatorial view of stochastic processes: White noise 
}

\author{A. Diaz-Ruelas}
 \altaffiliation[]{Max Planck Institute for the Physics of Complex Systems, Nöthnitzer Str. 38, 01187 Dresden.\\ Max Planck Institute for the Mathematics in the Sciences, Inselstr. 22, 04103 Leipzig.}

\date{\today}

\begin{abstract}

White noise is a fundamental and fairly well understood stochastic process that conforms the conceptual basis for many other processes, as well as for the modeling of time series. Here we push a fresh perspective toward white noise that, grounded on combinatorial considerations, contributes to give new interesting insights both for modelling and theoretical purposes. 
To this aim, we incorporate the ordinal pattern analysis approach which allows us to abstract a time series as a sequence of patterns and their associated permutations, and introduce a simple functional over permutations that partitions them into classes encoding their level of asymmetry.   
We compute the exact probability mass function (p.m.f.) of this functional over the symmetric group of degree $n$, thus providing the description for the case of an infinite white noise realization. This p.m.f. can be conveniently approximated by a continuous probability density from an exponential family, the Gaussian, hence providing natural sufficient statistics that render a convenient and simple statistical analysis through ordinal patterns. Such analysis is exemplified on experimental data for the spatial increments from tracks of gold nanoparticles in 3D diffusion. 

\end{abstract}

\maketitle


\section{\label{Sec:Intro} Introduction }

The incorporation of stochastic ingredients in models describing phenomena in all disciplines is now a standard in scientific practice. White noise is one of the most important of such stochastic ingredients. Although tools for identifying white and other types of noise exist \cite{Gelman2013,HolgerBook}, there is a permanent demand for reliable and robust statistical methods for analyzing data in order to distinguish noise and filter it from signals in experiments. Or in hypothesis tests, for assessing the plausibility of the outcome of an experiment being the result of randomness and not a significant, controllable effect. Due to its ubiquity in experiments and its mathematical simplicity, white noise is very often the most convenient stochastic component that adds realism to a dynamic model, commonly regarded as the noise polluting observations. It can be continuous or discrete both in time or in distribution, so it can be applied to many scenarios. It is a stationary and independent and identically distributed process, all relatively simple properties for a stochastic process. Here we present a combinatorial perspective to study white noise inspired in the concept of ordinal patterns. 
An ordinal pattern of length $n$ is the diagramatic representation of the inequality fulfilled by a subsequence of $n$ points $x_1,…,x_n$ in a time series $\{x_t\}_{t\in I}$. We discuss ordinal patterns in detail in Sec. \ref{Sec:OrdPatterns}. This concept was used in 2002 by Bandt and Pompe \cite{BandtPompe2002} for building a measure of complexity for time series named Permutation Entropy (PE).  PE has proven its value not only in applications, when used to analyze time series from a great variety of phenomena \cite{Zanin2012,Zanin2021}, but it is also of theoretical relevance since it is equivalent to the Kolmogorov-Sinai entropy for a large class of piece-wise continuous maps of the interval \cite{BandtKellerPompe2002,Misiurewicz2003}. The procedure for computing PE consists in first choosing the size $n$ for the window that will be used in the analysis, corresponding to the embedding dimension in a Takens embedding \cite{HolgerBook}. Then we slide the window through the series (equivalent to construct the lagged or delay vectors \cite{HolgerBook}), to find the frequencies $\pi_j, k=1,…,n!$ with which ordinal patterns occur. Then we compute the associated Shannon entropy $H_{\pi} = -\sum_j \pi_j \log \pi_j$. For white noise in the long time series limit all the patterns are equally likely, therefore their frequencies follow a discrete uniform probability mass function (p.m.f.) with support on the integers $1,…,n!$, which is equivalent to the probability measure of permutations over the symmetric group of degree $n$, $S_n$.  Despite its relevance and wide range of applications, there are few rigorous studies on the properties of PE for its use in statistical inference. To the best of the author’s knowledge, this is addressed only in works such as \cite{LittleKane2016}, where the authors investigate the expectation value and variance of PE for finite time series of white noise, and later the same authors address the effect of ordinal pattern selection on the variance of PE \cite{LittleKane2017}. 
In the customary PE approach every permutation is, in a sense, considered as a class, since the count of every single permutation is important. The effect of this is that the empirical distributions obtained from finite-length observations will be very sensitive to relatively minor changes in the proportions of each observed pattern \cite{LittleKane2016}. This lack of robustness represents a liability when trying to distinguish noise from structure. Another problem is the factorial growth of the number of classes, enlarging the discrete support correspondingly and making the analysis both impractical and meaningless for values of the embedding dimension beyond low or moderate $n$ ($\sim 10$), since the required length of a time series that will have a chance to display roughly one representative member of each class would be on the same order of $n!$ (already around 3 million observations for $n=10$).

In this work, we address these limitations by introducing a new statistic for permutations in Sec. \ref{Sec:AlphaTau}. This statistic is a functional over the symmetric group $S_n$, that can be interpreted as a measure of asymmetry for ordinal patterns. The functional divides the symmetric group into classes corresponding to a coarse notion of levels of overall increasing or decreasing behaviour of a pattern. In turn, this results on the transformation of the original discrete uniform probability measure of the patterns over $S_n$, into a new probability measure that concentrates around its expected value, as we show in Sec \ref{Sec:SuffStats}. This has practical and conceptual consequences, such as the ability of performing a suitably modified version of the ordinal pattern analysis for very large embedding dimensions, since now the number of classes of patterns to be tracked is reduced from $\mathcal{O}(n!)$  to merely $\mathcal{O}(n^2)$. The probability mass functions corresponding to our functional can be approximated by a Gaussian distribution, that is itself an exponential family. This guarantees the existence of natural sufficient statistics for the estimation of our statistic, as we explain in Sec \ref{Sec:SuffStats}. We open Sec. \ref{Sec:TSA_StatInference} with an illustration of our framework by analysing white noise from different source distributions and discussing its potential for distinguishing the deterministic signature of a chaotic orbit from a discrete map by inspection of the empirical p.m.f. obtained from our modified pattern analysis. Then we take experimental 3D tracks of gold nanoparticles and design a test for identifying the statistical independence among the spatial increments on each coordinate in a plane of observation. We finish by discussing our results and making additional remarks on the advantages and drawbacks of the presented framework.  

\section{Ordinal Patterns \protect\\ and their symmetries \label{Sec:OrdPatterns} }

A permutation $\tau$ is a bijection $\tau: S\rightarrow S$. If we take the set $S$ to be the sequence of integers $S=\{1,2,3,\ldots,n\}$, then $\tau(i)= \tau_i$  maps this sequence into itself $\tau_i = k, \  \ i,k = 1,2,3,\ldots,n $. Due to its bijective nature, we call the arrangement $\tau = \tau_1\tau_2\cdots\tau_n$ also a permutation of length $n$. Alternatively we can call it a word produced by $\tau$ on $n$ symbols. The set $S_n = \{\tau: \tau(i) = k, \ \ i,k = 1,2,3,\ldots,n \}$ together with the operation of functional composition is the symmetric group on $n$ symbols,  of cardinality $|S_n| = n!$ \cite{CombPerm_Bona2012}. We will denote the set of symbols as $[n]:= \{1,2,\ldots,n\}$ and an interval of symbols is denoted by $[i,j]\in[n]$. 

\noindent  From a simplified perspective, the elements $\tau \in S_n$ can be regarded all equivalent to each other a priori. Therefore, the corresponding discrete measure over $S_n$ would be $U(1,n!)$, assigning each permutation a weight of $1/n!$. Under this measure, permutations can be enumerated by lexicographic order, which ranks the words according to their size as integers. The lexicographic order in $S_3$ is shown in Table \ref{Tab:S_3_lexicographic} following the index $i$ along with other statistics for permutations, such as the inversion number, the major index and the runs (number of ascending sequences) \cite{CombPerm_Bona2012},\cite{Andrews2004}. In the last column, we include the corresponding values of the functional introduced in Sec. \ref{Sec:AlphaTau}. \\


\begin{table}[ht!]
\begin{center}
 \begin{tabular}{|c|c|c|c|c|c|c|}
 \hline
  $i$  & $\tau$ & sgn & inv & maj & runs & $\alpha(\tau)$\\ 
 \hline\hline
 1  & 123 & +1 & 0 & 0 & 1 & 2\\ 
 \hline
 2  & 132 & -1 & 1 & 2 & 2 & 1\\ 
 \hline
 3  & 213 & -1 & 1 & 1 & 2 & 1\\ 
 \hline
 4  & 231 & +1 & 2 & 2 & 2 & -1\\ 
 \hline
 5  & 312 & +1 & 2 & 1 & 2 & -1\\ 
 \hline
 6  & 321 & -1 & 3 & 3 & 3 & -2\\ 
 \hline
 
\end{tabular}
\end{center}
\caption{Sign, inversion number, major index, runs and the functional $\alpha(\tau)$ introduced in Sec.\ref{Sec:AlphaTau} for the permutations $\tau\in S_3$, listed in lexicographic order. }
\label{Tab:S_3_lexicographic}
\end{table}

Consider the lattice  $L_n  $ formed by the cartesian product $S_n \times S_n$. An \textit{ordinal pattern} \cite{BandtPompe2002}, that we will denote here as $d_\tau= \langle \tau(1),\tau(2),\ldots,\tau(n) \rangle$ or simply $d$ is a directed graph in $L_n$ that connects ordered pairs that are consecutive in their first index $(i,\tau(i))\rightarrow(i+1,\tau(i+1))$, $i=1,2,\ldots,n$, hence $d:S\times S \rightarrow S_n$. The angle brackets $\langle \cdot \rangle$ explicitly indicate that the elements of the $n$-tuple $(\tau(1),\tau(2),\ldots,\tau(n))$ are connected consecutively conforming a graph.

For instance, the graph $(1,1)\rightarrow(2,2)$, or $\langle 1,2 \rangle$ in the $L_2$ lattice is an ordinal pattern, since $\tau(1)=1,\tau(2)=2$, then $d_{\mathrm{id}}=\langle 1,2 \rangle$ corresponds to the identity permutation in $S_2$, $\tau_\mathrm{id} = 12$. The graph $(1,1)\rightarrow(2,1)=\langle 1,1 \rangle $ is not a valid diagram since the word $11$ is not a permutation, \textit{i.e.} $11 \notin S_2 = \{12,21\}$. The set of all diagrams is denoted by $D_n = \{d: d=\langle \tau(1),\tau(2),\ldots,\tau(n)  \rangle \}$.  Tables \ref{Tab:n=3_CharPerm} and \ref{Tab:n=4_CharPerm} display the permutations and their corresponding graph sets $D_3$, $D_4$, respectively.


\begin{table}[htb!]
\begin{center}
 \begin{tabular}{|c|c|c|c|c|c|c|c|c|}
 \hline
 $\tau$ & $\alpha(\tau) $  & pattern & $\tau$ & $\alpha(\tau) $  & pattern & $\tau$ & $\alpha(\tau) $  & pattern\\ 
 \hline\hline

 123 &  2 & \raisebox{-\totalheight}{\includegraphics[width=0.04\textwidth , height=6mm]{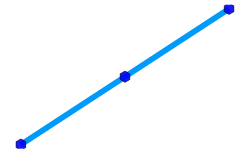}} &
 213 & 1 & \raisebox{-\totalheight}{\includegraphics[width=0.04\textwidth , height=6mm]{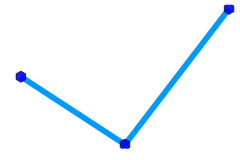}} & 
 312 & -1 &\raisebox{-\totalheight}{\includegraphics[width=0.04\textwidth , height=6mm]{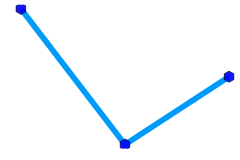}} \\
 
 \hline  
 
 132 &  1 & 
 \raisebox{-\totalheight}{\includegraphics[width=0.04\textwidth , height=6mm]{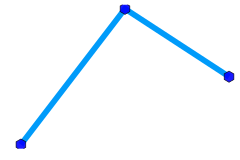}} & 
231 & -1 & \raisebox{-\totalheight}{\includegraphics[width=0.04\textwidth , height=6mm]{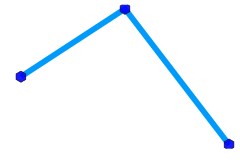}} & 
 321 & -2 &
 \raisebox{-\totalheight}{\includegraphics[width=0.04\textwidth , height=6mm]{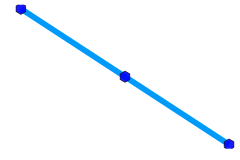}}\\ 

 \hline
\end{tabular}
\end{center}
\caption{Permutations $\tau$, $\alpha(\tau)$, and the associated ordinal patterns of length $n=3$.}\label{Tab:n=3_CharPerm}
\end{table}


\begin{table*}[t]
\caption{\label{Tab:n=4_CharPerm} Permutations $\tau$, the functional $\alpha(\tau)$ (defined in Sec. \ref{Sec:AlphaTau}), and the associated ordinal patterns of length $n=4$.  Reflection of the patterns on the first and second columns across a horizontal axis results in the diagrams in the pervious to last and last columns, respectively, with the same absolute values of the characteristic for the reflected diagrams/permutations, but negative sign. This is a basic property of $\alpha(\tau)$ that is also reflected statsistically.}
\begin{center}

 \begin{tabular}{|c|c|c|c|c|c|c|c|c|c|c|c|}
 \hline
 $\tau$ & $\alpha(\tau) $  & pattern & $\tau$ & $\alpha(\tau) $  & pattern & $\tau$ & $\alpha(\tau) $  & pattern & $\tau$ & $\alpha(\tau) $  & pattern \\ 
 \hline\hline 
 1234 &  4 & \raisebox{-\totalheight}{\includegraphics[width=0.05\textwidth, height=6mm]{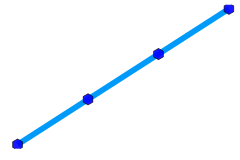}} & 
 2134 & 4 & \raisebox{-\totalheight}{\includegraphics[width=0.05\textwidth, height=6mm]{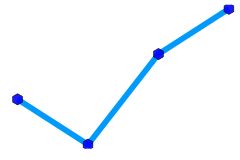}} & 
 3124 & 2 & \raisebox{-\totalheight}{\includegraphics[width=0.05\textwidth, height=6mm]{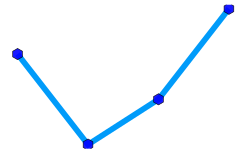}} &
 4123 & 0 & \raisebox{-\totalheight}{\includegraphics[width=0.05\textwidth, height=6mm]{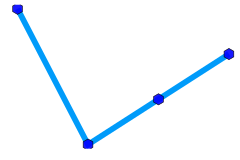}}   
 \\
 
 \hline
 
 1243 &  4 & \raisebox{-\totalheight}{\includegraphics[width=0.05\textwidth, height=6mm]{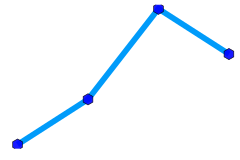}} & 
 
2143  & 4 & \raisebox{-\totalheight}{\includegraphics[width=0.05\textwidth, height=6mm]{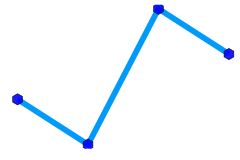}} & 
 
3142 & 2 & \raisebox{-\totalheight}{\includegraphics[width=0.05\textwidth, height=6mm]{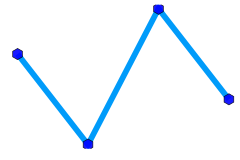}} & 
 
 4132 & 0 & \raisebox{-\totalheight}{\includegraphics[width=0.05\textwidth, height=6mm]{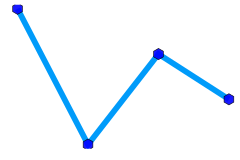}} \\ 
 
 \hline

 1324 & 2 & \raisebox{-\totalheight}{\includegraphics[width=0.05\textwidth, height=6mm]{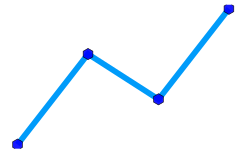}}& 
 2314 & 0 & \raisebox{-\totalheight}{\includegraphics[width=0.05\textwidth, height=6mm]{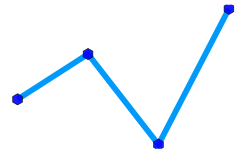}}& 
 3214 & 0 & \raisebox{-\totalheight}{\includegraphics[width=0.05\textwidth, height=6mm]{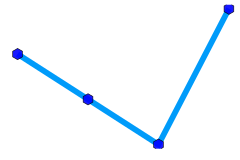}}& 
 4213 & -2 & \raisebox{-\totalheight}{\includegraphics[width=0.05\textwidth, height=6mm]{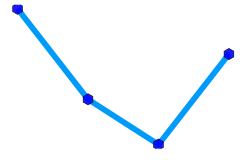}}\\
 
 \hline

 1342 & 2 & \raisebox{-\totalheight}{\includegraphics[width=0.05\textwidth, height=6mm]{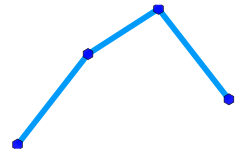}}& 
 2341 & 0 & \raisebox{-\totalheight}{\includegraphics[width=0.05\textwidth, height=6mm]{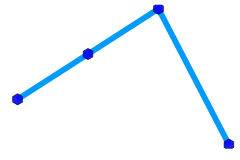}}& 
 3241 & 0 & \raisebox{-\totalheight}{\includegraphics[width=0.05\textwidth, height=6mm]{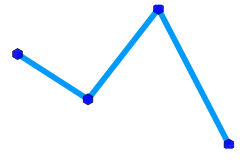}}& 
 4231 & -2 &  \raisebox{-\totalheight}{\includegraphics[width=0.05\textwidth, height=6mm]{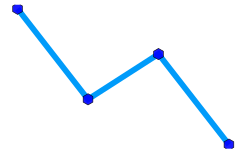}}\\
 
 \hline

 1423 & 0 & \raisebox{-\totalheight}{\includegraphics[width=0.05\textwidth, height=6mm]{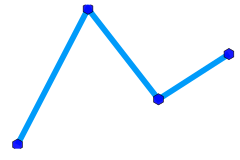}}& 
 2413 & -2 & \raisebox{-\totalheight}{\includegraphics[width=0.05\textwidth, height=6mm]{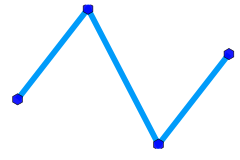}}& 
 3412 & -4 & \raisebox{-\totalheight}{\includegraphics[width=0.05\textwidth, height=6mm]{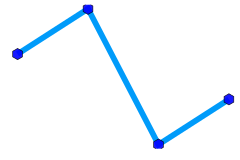}}& 
 4312 & -4 & \raisebox{-\totalheight}{\includegraphics[width=0.05\textwidth, height=6mm]{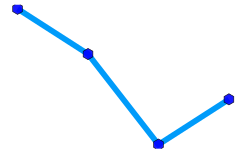}}\\ 
  
 \hline
 
 1432 & 0 & \raisebox{-\totalheight}{\includegraphics[width=0.05\textwidth, height=6mm]{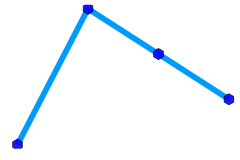}}& 
 2431 & -2 & \raisebox{-\totalheight}{\includegraphics[width=0.05\textwidth, height=6mm]{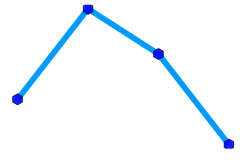}}& 
 3421 & -4 & \raisebox{-\totalheight}{\includegraphics[width=0.05\textwidth, height=6mm]{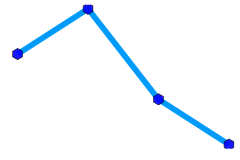}}& 
 4321 & -4 &  \raisebox{-\totalheight}{\includegraphics[width=0.05\textwidth, height=6mm]{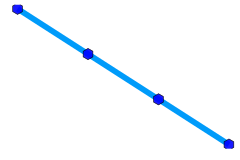}}\\ 
 
 \hline
\end{tabular}
\end{center}
\end{table*}

\subsection{Reflections of diagrams}\label{sSec:TransfDiag}

It is instructive to explore the symmetries of the ordinal patterns, since they are reflected in the statistical properties of the permutations as we explain in Sec. \ref{Sec:AlphaTau} for inspecting the properties of the distribution of the statistic introduced there. \\

\subsubsection*{Vertical reflections}

A vertical reflection $v(d)$ consists on flipping a diagram along a vertical axis and relabeling the nodes (ordered pairs) accordingly. It is a bijection $h:D\rightarrow D$ that acts on ordered pairs as 

\begin{equation}
v(i,\tau_i)=(v(i),\tau_i),
\end{equation}

\begin{figure}[h!]
\centering
\includegraphics[width=0.55\columnwidth]{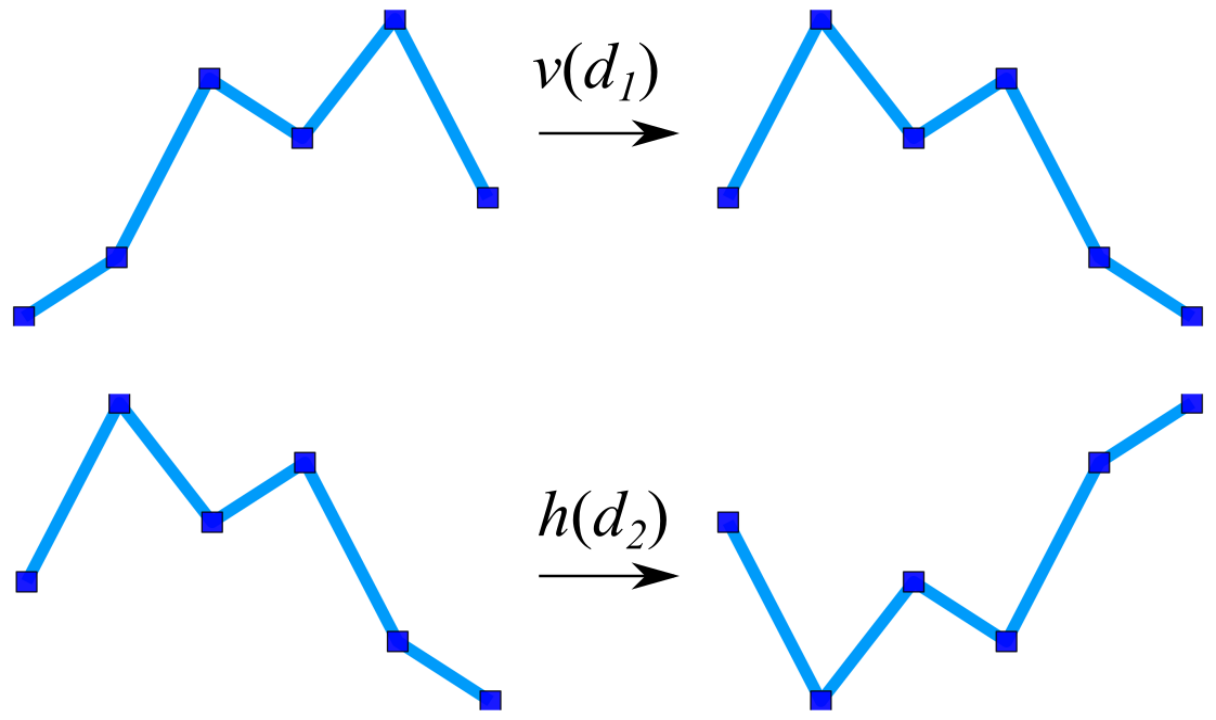}
\caption{Effect of vertical and horizontal reflections on diagrams $d_1 = \langle 1,2,5,4,6,3 \rangle$ and $d_2 = \langle 3,6,4,5,2,1 \rangle$ respectively. (see Sec. \ref{Sec:AlphaTau}).} \label{Fig:DiagramReflections}
\end{figure}

\noindent \textit{i.e.} it leaves the second coordinate invariant. The action of $v$ over $[n]$ is explicit

\begin{equation}
v(1)=n, \ \ v(2) = n-1,  \ldots, \ \ v(n-1) = 2, \ \ v(n) = 1 , \label{Eq:v_explicit_action}
\end{equation} 

\noindent meaning that $v$ simply mirrors the symbols along a vertical axis of reflection to the right of the permutation, as illustrated in Fig. \ref{Fig:DiagramReflections}. For even permutation length $n$, the effect of $v(\tau)$ as a reflection with respect to an external vertical axis, is the same as an internal reflection of the symbols of the word $v(\tau_1\tau_2\cdots\tau_{n-1}\tau_n) = \tau_n\tau_{n-1}\cdots\tau_2\tau_1$ along an internal axis splitting the permutation in equal parts of $\frac{n}{2}$ symbols. For odd $n$, the permutation is split into equal parts of $m \equiv \left \lfloor{\frac{n}{2}}\right \rfloor $ but the symbol $\tau_{m+1}$ is left invariant since it becomes the internal axis of reflection itself. 
As a simple illustration, let us take a diagram of length $n=6$, as shown schematically in Fig. \ref{Fig:DiagramReflections}(a). Ranking the points from smallest to largest gives the \textit{ranking permutation} $\tau=\tau_1\tau_2\cdots\tau_n$, corresponding to the permuted indices of the vector of ranks:  $\mathrm{r}(x_{t_1},x_{t_2}\ldots,x_{t_n}) = (x_{\tau_1},x_{\tau_2}\cdots,x_{\tau_n} )$ relative to the original positions $t_1,t_2,\ldots t_n$. Hence, in the example of Fig. \ref{Fig:DiagramReflections}(a), $v(\langle 1,2,5:4,6,3\rangle)=\langle 3,6,4:5,2,1 \rangle$, where we put double dots instead of a comma in the middle only to highlight the internal axis of symmetry of the diagram.

\subsubsection*{Horizontal reflections}

A horizontal reflection $h(d)$ consists on flipping a diagram along a horizontal axis. Its action over ordered pairs is  

\begin{equation}
h(i,\tau(i)) = (i,h(\tau_i)), 
\end{equation}

\noindent meaning that $h$ acts directly on the permutation symbols. Its explicit action corresponds to the involution

\begin{equation}
 \langle\tau_1,\tau_2,\cdots,\tau_n\rangle \rightarrow \langle n+1-\tau_1,n+1-\tau_2,\cdots,n+1-\tau_n\rangle, \label{Eq:h_explicit_action_Involution}
\end{equation}




\section{A functional over permutations \label{Sec:AlphaTau}}

All of the permutation statistics displayed in Table \ref{Sec:Intro} reflect different symmetries in $S_n$. The sign of the permutation is arguably the most basic statistic, telling explicitly whether the number of flips of symbols in a permutation $\tau$ relative to the identity $\tau_{\mathrm{id}}=123\cdots n$ is even ($\sgn(\tau)=+1$) or odd ($\sgn(\tau)=-1$). The inversion number explicitly counts the pairs for which the symbol in position $i$ is larger than the one at $i+1$. Summing up the indices of the larger symbols in each inversion yields the major index $\mathrm{maj}(\tau)=\sum_{\tau(i)>\tau(i+1)} i$. \\

\noindent Below we introduce a new statistic of permutations that accounts for an imbalance in weight when considering the symbols $[n]$ conforming every permutation $\tau \in S_n$ as a collection of weights. The objective of defining and characterizing this new statistic is twofold: On one hand, we have the intrinsic interest on new insights on the study of the symmetric group and the implications of these discoveries on other areas such as dynamical systems and stochastic processes. On the other hand, there is also a practical interest on the analysis of time series, specifically in connection with ordinal pattern analysis. 

The construction of the set of ordinal patterns from a realization of any process $X_t$, consists on mapping portions of that process time series $\{ x_t \}_{t \in I}$ (with  $ I = [L]$ an index set, usually $[N]={1,2,\ldots,L}$), to the diagrams in Sec.\ref{Sec:OrdPatterns} by sliding a window of size $n$ over $\{ x_t \}_{t \in I}$. We can inspect larger portions of the series without increasing $n$ by skipping (lagging) a number of $l-1$ points after every choice on each window, so the number $l$ is correspondingly known as lag. For instance, a lag of $l=1$ means that we do not skip any point and we slide the window without gaps. The collection of windows of $n$ points are called delay or lagged vectors, and the process of construction of these vectors is known as an embedding \cite{HolgerBook}. Correspondingly, the natural number $n$ indicating the window size is called embedding dimension. This terminology comes from the embedding theorems that form the basis of the state space reconstruction methods from scalar time series, known in general as time delay embeddings \cite{HolgerBook}. Hence, for the ordinal pattern analysis, we first make the embedding of the time series for a chosen dimension $n$ and lag $l$. This yields a total of $N = L - (n-1)l$ lagged vectors. Then we rank the  amplitudes $\{x_1,x_2,\ldots,x_n \}$ on every lagged vector according to their magnitude, thus obtaining the ranked vectors $\{x_{\tau(1)},x_{\tau(2)},\ldots,x_{\tau(n)}\}$. The sequence of indices in these ranked vectors are the ranking permutations $\tau = \tau_1\tau_2\cdots\tau_n$, where $\tau_i=\tau(i)$. Finally, these permutations have associated ordinal patterns as we saw in Sec. \ref{Sec:OrdPatterns}. In this way the local information on the relative ordering is preserved, and as a consequence also the relevant information on the correlation structure of the series. 
 However, a simple but careful inspection of this mapping makes evident that the relative ordering is not the only information preserved. In fact, if we interpret the ranks as abstract weights assigned to the amplitudes of the process, then it is clear that also information on the overall variation within portions of these windows of $n$ data points is preserved. Questions such as the weight accumulated in sub-intervals within each bin of size $n$ arise naturally from this observation. An equally natural choice of sub-interval for analysis within the $n$-window is half the bin, so as to build a measure to estimate the tendency of the process to have local increasing, decreasing or close to constant behavior. This is in close analogy to the concept of derivative, but getting rid of the information on the specific amplitudes that the process takes. In order to study this asymmetry on the total weight concentrated on each half of a length $n$ window, 
  we provide the following

\begin{quote}
\textbf{Definition 1} Functional $\alpha(\tau)$. \textit{For every permutation $\tau\in S_n$, $n$ a non-negative integer, define the functional}
\begin{equation}
\alpha(\tau) = 
\begin{cases}
\sum_{i= m+2}^{n}\tau(i) - \sum_{i=1}^{ m }\tau(i), \ n \ \mathrm{odd} \\
\sum_{i=m+1}^{n}\tau(i) - \sum_{i= 1 }^{m}\tau(i) ,\ n \ \mathrm{even },
\end{cases}
\label{Eq:AlphaTau_Def}
\end{equation}
\textit{where $m\equiv\lfloor{ \frac{n}{2}\rfloor}$.} 
\end{quote}

 The functional thus defined is invariant under a shift of the sequence $[n]$ by an integer. This transformation is also invariant under monotonic transformations of the process $X_t$, since the relative ordering is not affected. The case for odd $n$ gives mixed statistical behavior, since the middle permutation symbol $\tau_{m+1}$ is ignored in the computation of $\alpha(\tau)$, implying that for a given permutation lenght $n$, when the ignored symbol happens to be $\tau_{m+1} = n$, there will be $(n-1)!$ permutations that display the same statistics as in the full problem for permutation lenght $n-1$. By the invariance of $\alpha(\tau)$ under a shift of $\{1,2,\ldots, n\}$ by an integer, when the middle symbol is $\tau_{m+1}=1$ we have the same situation as before. Other ignored symbols produce different and more complicated effects. Therefore, we shall limit here to the case for even $n$, which reduces Eq. (\ref{Eq:AlphaTau_Def}) to 

\begin{equation}
\alpha(\tau) =  \sum_{k=1}^{m} \left[\tau\left(k+m\right)-\tau\left(m-k+1\right)\right],  \ \ m=\frac{n}{2}.
\label{Eq:AlphaTau_Def_even_n}
\end{equation}

\noindent Hence, $\alpha(\tau)$ splits $\tau$ into equally sized intervals $L=[\tau(1),\tau(m)]$ and $R=[\tau(m+1),\tau(n)]$ with partial sums $s_{l} = \sum_{i=1}^{m}\tau(i)$ and $s_{r} = \sum_{i=m+1}^{n}\tau(i)$, respectively. We get $\alpha(\tau)$ by summing up the symbols on each half, and subtracting: $\alpha = s_{r} - s_{l}$. Notice that  $s_{l,\mathrm{min}}=\frac{m(m+1)}{2}$, and $s_{l,\mathrm{max}} = \frac{n(n+1)}{2} - \frac{m(m+1)}{2} = \frac{3m^2+m}{2}$. This means that $\alpha_{\mathrm{max}}\equiv \alpha_M = m^2 $, as it is the case, for instance, for the identity permutation $\alpha(\tau_\mathrm{id})=\alpha_M$. By symmetry, using the reflections in Eqs. (\ref{Eq:v_explicit_action}) and (\ref{Eq:h_explicit_action_Involution}) the minimum value of $\alpha$ is $\alpha(h(\tau_\mathrm{id}))=\alpha(v(\tau_\mathrm{id}))=-\alpha_M$. This means that $\alpha(\tau)$ is a bounded variable for finite $n$. 

 The definition of $\alpha(\tau)$ allows for an obvious source of degeneracy or multiplicity, since the order of the summands on each half $s_l,s_r$ is irrelevant. Let us denote the number of permutations that share the same value of $\alpha(\tau)$ by $\phi(\alpha)$. For every distinct value of $\alpha(\tau)$, there are (by shuffling the terms on each sum $s_l, s_r$) \textit{at least} $|L|\cdot|R|=\left(m!\right)^2 \equiv \eta$ reorderings that share the same value, hence $\phi(\alpha)\geq \eta$. For instance, if we consider $\tau_\mathrm{id}$ we get $\alpha(\tau_\mathrm{id})=\alpha_M$ as above, which has the least degeneracy $\phi(\alpha_M)= \eta$. The next possible value of $\alpha(\tau)$ can be obtained from $\tau_\mathrm{id}=123\cdots n$ by switching $m$, the largest symbol in $L$, with the smallest symbol in $R$, which is $m+1$. This clearly  increases the value of the left partial sum by one $s_l\rightarrow s_l+1$ while decreasing the right partial sum by one unit $s_r\rightarrow s_r-1$, with the effect of decreasing $\alpha(\tau_\mathrm{id})$ by two units $\alpha(\tau_\mathrm{id}) \rightarrow \alpha(\tau_\mathrm{id})-2$, then the next value is $\alpha'=\alpha_M-2$. Proceeding in this way, we get the set of all possible $\alpha$-values over $S_n$ (for even $n$)

\begin{equation}
A_n = \{-\alpha_M,-\alpha_M+2,-\alpha_M+4,\ldots,\alpha_M-4,\alpha_M-2,\alpha_M \}
\label{Eq:alpha-support}
\end{equation}

\noindent whose size is $|A_n| = m^2+1$. 
Notice that if $m^2 \equiv 1 \mod(2)$, then $\min_{\tau}\{|\alpha(\tau)|: \tau\in S_n \} = 1$ and $\min_{\tau}\{|\alpha(\tau)|: \tau\in S_n \} = 0$ for $m^2 \equiv 0 \mod(2)$, a difference that becomes irrelevant as $n\rightarrow \infty$. It is not difficult to see that the degeneracies come in multiples of $\eta$, meaning that $\phi(\alpha=\alpha_i)=a_i\eta$, $\alpha_i=\pm \alpha_M \mp 2i$, $i=0,1,2,\ldots,m^2$. The coefficients $a_i(n)$ for $i \in [1,\lfloor \frac{m^2}{2} \rfloor + 1]$ form sequences

\begin{eqnarray}
a_i(2) &=& 1,1 \nonumber \\ 
a_i(4) &=& 1,1,2 \nonumber \\  
a_i(6) &=& 1,1,2,3,3 \nonumber\\ 
a_i(8) &=& 1,1,2,3,5,5,7,7,8 \nonumber \\  
a_i(10) &=& 1,1,2,3,5,7,9,11,14,16,18,19,20 \nonumber \\ 
a_i(12) &=& 1,1,2,3,5,7,11,13,18,22,28,32,39,42,48, \nonumber \\ 
        & & 51,55,55,58 \nonumber \\ 
a_i(14) &=& 1,1,2,3,5,7,11,15,20,26,34,42,53,63,75, \nonumber \\ 
          & & 87,100,112,125,136,146,155,162,166,169 \nonumber \\ 
          &\vdots & 
\label{Eq:a_ni_sequence}
\end{eqnarray}

\noindent that are mirrored for $i\in [\lfloor \frac{m^2}{2} \rfloor + 2,n]$. The sequences in Eq. (\ref{Eq:a_ni_sequence}) were found by direct numerical computation, but in the following we obtain them analytically from the observations in the construction of the set $A_n$ and the definition in Eq (\ref{Eq:AlphaTau_Def_even_n}) for $\alpha(\tau)$.

It is clear that $\sum_i \phi(\alpha_i) = \eta \sum_i a_i = n!$, therefore,

\begin{equation}
\sum_{i = 0}^{m^2} a_i  = {{n}\choose{m}}, \ \ m=\frac{n}{2}.
\label{Eq:ai_series}
\end{equation}

\noindent Thus, the central binomial coefficient ${{n}\choose{m}}$ counts the number of ways of computing $\alpha(\tau)$ in a nontrivial way. This is because ${{n}\choose{m}}$ also counts the number of different ways to assign a $+$ sign to $m=n/2$  out of $n$ symbols and a $-$ sign to the rest of $n-m=n/2$ symbols, which is precisely the problem of computing $\alpha(\tau)$ without counting the shuffling of the terms in the sums. 
Yet, Eq. (\ref{Eq:ai_series}) tells us the total number of different ways to compute $\alpha(\tau)$, not the value of the individual coefficients $a_i$. To find the nontrivial multiplicities $a_i$, we notice the equivalence between our problem and the combinatorial problem of sums of partitions of sets. The multiplicities $a_i$ correspond to the number of permutations $\tau \in S_n$ such that $\alpha(\tau)=a_i$. This is equivalent to the problem of finding the number of sets made of $m=n/2$ elements out of $n$ total elements, such that the sum of the $m$-element set is fixed. This, in turn, fixes the sum of the $n-m$ remaining elements. These sums are clearly $s_l,s_r$ discussed before. Since fixing either of the sums fixes the value of $\alpha(\tau)$, we have thus just rephrased the computation of our functional as a combinatorial problem which has an elegant solution, stated in the form of the following

\begin{quote}
\textbf{Theorem} (Bóna 2012 \cite{CombPerm_Bona2012}) \textit{Let $n$ and $k$ be fixed non-negative integers so that $k \leq n$. Let $b_i$ denote the number of $k$-element subsets of $[n]$ whose elements have sum ${k+1 \choose 2}+i$, that is, $i$ larger than the minimum. Then we have}
\begin{equation}
{\mathbf{n} \brack \mathbf{k}} := \sum_{i=0}^{k(n-k)}b_i q^i.	
\label{Eq:GBinom_GenFun_partitions}
\end{equation}
\textit{In other words, ${\mathbf{n} \brack \mathbf{k}}$ is the ordinary generating function of the $k$-element subsets of $[n]$ according to the sum of their elements.} 
\end{quote}

\noindent  The object in Eq. (\ref{Eq:GBinom_GenFun_partitions}) belongs to a special kind of polynomials known as \textit{Gaussian polynomials} or Gaussian binomial coefficients \cite{CombPerm_Bona2012}, regarded as a generalization of the binomial coefficients and defined as

\begin{equation}
{\mathbf{n} \brack \mathbf{k}} = \frac{[\mathbf{n}]!}{[\mathbf{n-k}]![\mathbf{k}]!} = \frac{(1-q^n)(1-q^{n-1})\cdots(1-q^{n-k+1})}{(1-q)(1-q^2)\cdots(1-q^k)}
\label{Eq:Def_GaussianBinomial_Bona}
\end{equation}

\noindent which are polynomials of the form in Eq. (\ref{Eq:GBinom_GenFun_partitions}) whose degree is $k(n-k)$ and with symmetric coefficients $b_i=b_{k(n-k)-i}$. The symbol $[\mathbf{k}]$ it is defined as

\begin{equation}
[\mathbf{k}] = \frac{1-q^k}{1-q} = 1 + q + q^2 + \cdots + q^{k-1}, \ \ q \neq 1,
\label{Eq:q-analog_k}
\end{equation}

\noindent also a polynomial, which reduces to $[\mathbf{k}]=k$ in the limit $q\rightarrow 1$. From the same limit, we also recover ${n \choose k}$ from Eq. (\ref{Eq:Def_GaussianBinomial_Bona}). The polynomial $[\mathbf{k}]$ in $q$  is called a $q$-analog of $k$, a natural choice for generalizing a representation of nonnegative integers $k$ parametrized by $q$. The generalization of the factorial as a $q$-analog is the  $q$-factorial

\begin{eqnarray}
[\mathbf{k}]! &=& [\mathbf{1}][\mathbf{2}][\mathbf{3}]\cdots[\mathbf{k}] \nonumber \\
	   &=& (1+q)(1+q+q^2)\cdots(1+q+q^2\cdots +q^{k-1}). \nonumber \\ 
	   \label{Eq:q-factorial}
\end{eqnarray}

\noindent Our main result, the computation of the numbers $a_i(n)$ in Eq. (\ref{Eq:a_ni_sequence}), corresponding to the nontrivial contribution to $\phi(\alpha_i)=a_i\eta$ reduces to a corollary of the Theorem in Eq. (\ref{Eq:GBinom_GenFun_partitions}) for the particular choice $k=m=n/2$. In fact, using $k=m=n/2$ we recover a generating function for our sequence $a_i$ 

\begin{equation}
G_n(q) = \sum_{i=0}^{m^2}a_i q^i, \ \ a_i=a_{m^2-i} > 0, \ \ q\neq 1,  
\end{equation}

\noindent with $G_n(1)={n \choose m}$. Therefore

\begin{equation}
a_i = [q^i]{\mathbf{n} \brack \mathbf{m}} = [q^i]G_n(q)
\label{Eq:ai_GPoly}
\end{equation}
 
\noindent  where the notation $[q^i]P(q)$ indicates the coefficient of the $i$-th power of the polynomial $P(q)$.
We can compute statistical properties of a sequence from its generating function with relative ease \cite{generatingfunctionology}. If we sample a permutation $\tau \in S_n$ at random and assign the random variable $\chi$ to the associated value of $\alpha(\tau)$, the probability of observing the value $\alpha_i$ is given by the ratio 

\begin{equation}
P_{\chi_n}(\alpha_i) = \frac{a_i}{\sum_{j=0}^{m^2}a_j} = \frac{[q^i]{\mathbf{n} \brack \mathbf{m}}}{{n \choose m}},
\label{Eq:pmf_chi}
\end{equation}

\noindent hence, $P_{\chi_n}(\alpha)$ is a probability mass function with support in $A_n$ (Eq. \ref{Eq:alpha-support}) and Eq. (\ref{Eq:ai_series}) is the normalization condition $\sum_i P_\chi(\alpha_i)=1$. The associated probability generating function (p.g.f.) is 

\begin{eqnarray}
G_{\chi_n}(q) &=& \sum_{0\leq i \leq m^2} p_i q^i \\
		  &=& \frac{G_n(q)}{G_n(1)} \\
		  &=& \frac{{\mathbf{n} \brack \mathbf{m}}}{{n \choose m}}
		  \label{Eq:pgf_chi}
\end{eqnarray}

\noindent where $p_i = P_{\chi_n}(\alpha_i)$. 
From the p.g.f in Eq. (\ref{Eq:pgf_chi}) we can, at least in principle, compute any desired moment of $\chi$ by differentiation of $G_\chi(q)$ with respect to $q$. However, the derivatives $G'_\chi(q)$ are more cumbesome than illuminating and we will not present them here. 

By applying the functional $\alpha(\tau)$ defined in Eq. (\ref{Eq:AlphaTau_Def_even_n}) to the set of observed permutations obtained from embedding a stochastic process $X_t$ as explained previously, we get an associated stochastic process in terms of $\alpha$. The functional $\alpha(\tau)$ can be applied to any process, but here we limit to its use for the analysis of white noise processes.

A white noise process $X_t$ is a continuous or discrete time stochastic process with the following properties
\begin{eqnarray}
E[X_t] = 0,\ \mathrm{and} \ \mathrm{Var}[X_t] = \sigma^2, \ \mathrm{for \ all \ } t \in \Omega \\
E[X_t X_s] = \sigma^2\delta(t-s), \ \mathrm{for \ all} \ s,t \in \Omega 
\end{eqnarray}
where $0<\sigma^2<\infty$, $\Omega$ is the set in which $s,t$ take values, for instance $\Omega=\mathbb{R}$ for a continuous process. From these properties, the zero-mean and finite variance conditions are irrelevant for the application of $\alpha(\tau)$, since they merely are information about the scale of the process, to which our functional is blind. 

The values of $\alpha$ over an embedding in a process are well defined, so long as the source process $X_t$ has a continuous support (later we will see that this condition can be sometimes relaxed), thus having a zero probability of observing repeated amplitudes of the process $X_t$. This means that the ranking permutations are well defined. Of course, real observations have a finite resolution, but even considering this, repetition of amplitudes from a stochastic process with continuous support would be expected to be highly unlikely at least in a finite time. Considering the former facts, we arrive at the following

\begin{quote}
\textbf{Definition 2} Induced $\chi$-process and $\alpha$-series. \textit{For every choice of non-negative integers $n$, $m$ and $l$ (with $n=2m$) and every white noise process $X_t$, the functional in Eq. (\ref{Eq:AlphaTau_Def_even_n}) induces the discrete-time process $\chi_k$ with discrete support $A_n$ (\ref{Eq:alpha-support}). Correspondingly, we denote a realization or time series of the process $\chi_k$ as}
\begin{equation}
\{\alpha_k \}_{k\in [N]},
    \label{Eq:AlphaProcess_Def}
\end{equation}
\textit{where $[N]=\{1,2,\ldots,N\}$, $L$ is the length of an observed realization of $X_t$, and $N = L -(n-1)l$, is the number of delay vectors in an embedding with fixed values $n$ and $l$.}
\end{quote}

\begin{figure}[t!]
\centering
\includegraphics[width=0.49\textwidth]{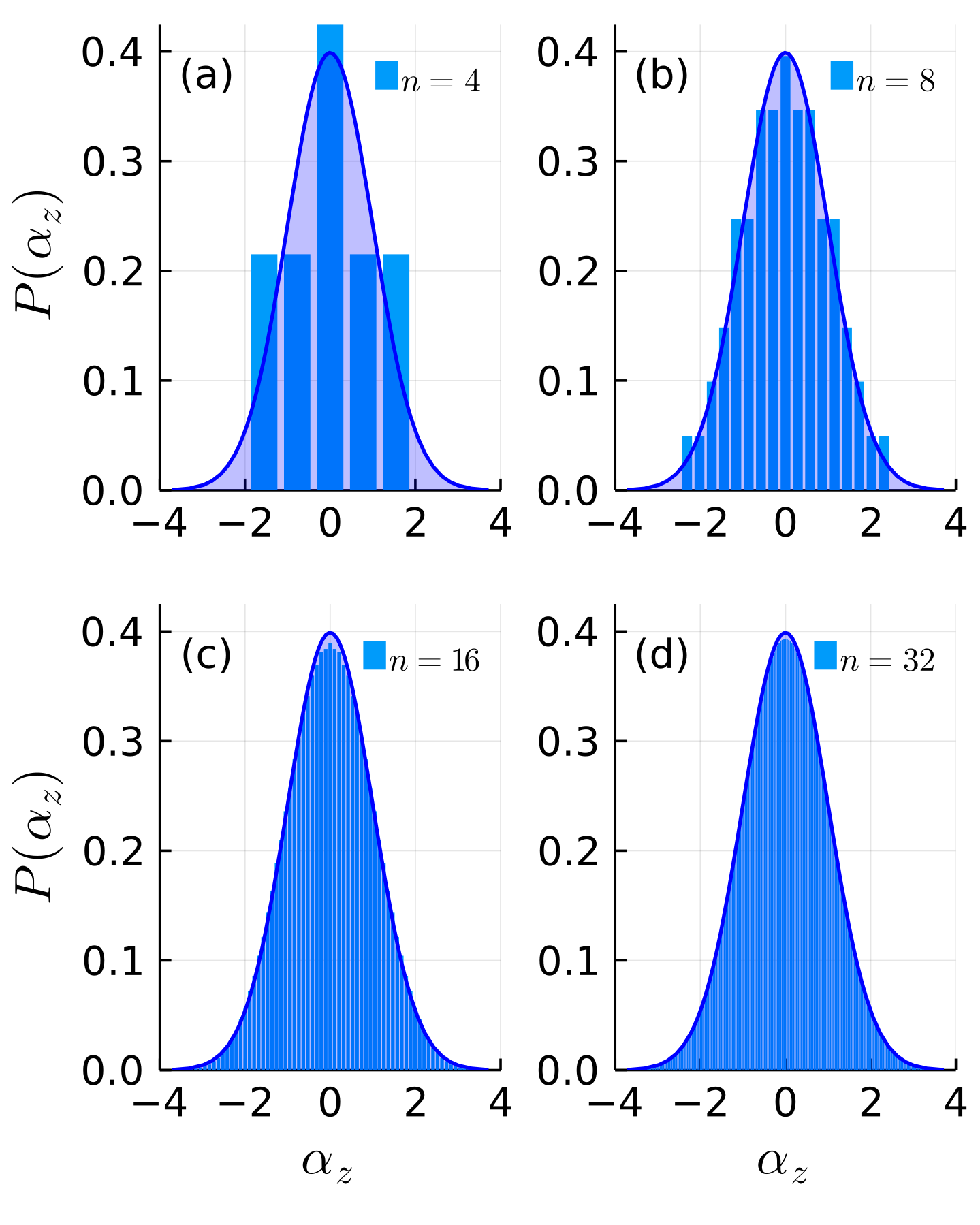}
\caption{Standardized probability mass function $P(\alpha_z)$ for $n=4,8,16,32$ represented by blue bars in panels (a),(b),(c) and (d) respectively, computed from Eq. \ref{Eq:ai_GPoly} and (\ref{Eq:pgf_chi}) using a recursive relation between Gaussian binomial coefficients. The continuous blue curve on each panel is a superimposed standard Gaussian ($\mu=\mu_\chi=0$, $\sigma=1$).} \label{Fig:pmf_n_4_8_16_32}
\end{figure}

In general, for arbitrary values of $n$ and $l$, the process $\chi_k$ and, consequently, $\{ \alpha_k \}_{k\in [N]}$ are correlated due to the overlap of the embedding vectors, that is in turn reflected in a sequence of permutations that are not independent. Nevertheless, the process could become effectively uncorrelated for some values of $l$ or at large values of both $n$ and $l$. 
In order to visualize how the p.m.f $P_{\chi_n}(\alpha_i)$ changes as $n$ increases, let us plot Eq. (\ref{Eq:pmf_chi}) for different $n$ values. To facilitate comparisons, we make use of the standardized variable $Z_\alpha = (\chi - \mu_\chi)/\sigma_\chi$. Since $\mu_\chi=0$, we have the simple quotient

\begin{equation}
    Z_\alpha = \frac{\chi}{\sigma_\chi}.
    \label{Eq:StandarsizedAlpha}
\end{equation}

\noindent we introduce the notation $Z_\alpha$ for referring to the standardized $\chi$ random variable. We will denote its realizations by $\alpha_z$ and the corresponding p.m.f. will be $P_{Z_\alpha}(\alpha_z)$ or simply $P(\alpha_z)$, with moments $\mu_{Z_\alpha}\equiv E[Z_\alpha]$, $\sigma_{Z_\alpha}\equiv\mathrm{Var}[Z_\alpha]$, and associated $\alpha$-process $\{ \alpha_{z,k} \}_{k\in [N]} $ (see (\ref{Eq:AlphaProcess_Def})). We plot $P(\alpha_z)$ for $n=4,8,16,32$ in Fig. \ref{Fig:pmf_n_4_8_16_32}. 

Let us come back briefly to the reflections of diagrams introduced in Sec. \ref{Sec:OrdPatterns} to add understanding to $P_\chi(\alpha)$. As illustrated in Figs. \ref{Fig:DiagramReflections} and \ref{Fig:DiagramReflections}, reflecting the diagrams change the corresponding values of $\alpha$ up to sign only. This means that for every $d\in D_n$ the compositions of $h$ and $v$ fulfill

\begin{equation}
\alpha[(h \circ v)(d)] = \alpha[(v \circ h)(d)],
\label{Eq:Compositions_h_v}
\end{equation}
 
\noindent and we will denote them simply by $hv(\cdot)$ and $vh(\cdot)$ in the following. Since in general $hv(d)\neq vh(d)$, this is the source of the degeneracy $\phi(\alpha)$ that is not accounted for by simple permutation of the terms of the partial sums $s_l, s_r$. An illustration of Eq. (\ref{Eq:Compositions_h_v}) is seen in Fig. \ref{Fig:DiagramReflections}. We get diagrams that are different in a nontrivial permutational way but with the same value of $\alpha(\tau)$ via the composition $hv$, as illustrated there by going from $d_1=\langle 1,2,5,4,6,3 \rangle$ to $d_{2} = \langle 4,1,3,2,5,6 \rangle$ 

\begin{eqnarray*}
\alpha(d_\tau) &=& -\alpha[v(d_1)]  \\
			   &=& \alpha[hv(d_1)] \\
			   &=& \alpha[d_{2}]\\
			   &=& 5 \\
\end{eqnarray*}

\noindent The symmetry of the coefficients  $a_i$ is thus equivalent to the reflection symmetry in Eq. (\ref{Eq:Compositions_h_v}).

\section{Continuous approximations and Sufficient statistics \label{Sec:SuffStats}}

By direct computation of the variance $\sigma_{\chi_n}^2=\sum_{0 \leq i \leq m^2 }\alpha_i^2 p_i$ for succesive $n$ values we arrive at the formula 

\begin{equation}
\sigma_\chi^2 = \frac{m^2(2m+1)}{3}
\label{Eq:VarChi_exact_formula}
\end{equation}

which for $m\rightarrow\infty$ becomes

\begin{equation}
\sigma_\chi^2 = \frac{2}{3}m^{3}
\label{Eq:VarChi_asymptotic}
\end{equation}

\noindent As previously noticed, the probability mass function $P_{\chi}(\alpha)$   converges to a Gaussian as $n$ increases. However, as can be noticed from Fig. \ref{Fig:pmf_n_4_8_16_32}  for low values of $n$ the shape of the distribution is different from a Gaussian, specially around its center. We discuss this case in Appendix \ref{App:A}.


The one-parameter exponential or Darmois-Koopman-Pitman families such as the Normal distribution arise naturally from the optimization of the Shannon entropy of $P(\chi)$ under normalization, first  and second moment $\sigma_\chi^2 = \frac{2}{3} m^{3}$ constraints \cite{Cover2005}  

\begin{eqnarray}
L[p] &=& -\sum_{i}p_i \log p_i -\beta_0(\sum_i p_i -1)  \\ \nonumber
	 & & -\beta_1 (\sum_{i}\alpha_i p_i -\mu_\chi) - \beta_2 (\sum_i\alpha_i^2p_i -\sigma_\chi^2)
	 \label{Eq:EntropyLagrangian}
\end{eqnarray}
 
optimization yields $p_i = \exp(1-\beta_0 - \beta_1\alpha_i-\beta_2\alpha_i^2)$. Using the normalization condition, together with $\mu_\chi=0$ we get

\begin{eqnarray}
p_i &=& \frac{e^{-\beta\alpha_i^2}}{\sum_j e^{-\beta\alpha_j^2}} \\  
	&=& \frac{a_i}{\sum_j a_j}  \nonumber 
	\label{Eq:pi_from_Lagrangian}
\end{eqnarray}

\noindent with $\beta=\beta_2$. We find $\beta$ by making a direct identification with the Gaussian p.d.f.   
\begin{equation}
f(x;\mu,\sigma) = \frac{1}{\sigma\sqrt{2\pi}}e^{\frac{(x-\mu)^2}{2\sigma^2}}
\label{Eq:GaussianPDF}
\end{equation}

\noindent with $\sigma = \sigma_{\chi_n}$ in formula (\ref{Eq:VarChi_asymptotic}), $\mu=\mu_{\chi_n}=0$ and $\sum_j P_{\chi_n}(\alpha_j)=1$ in order to get the correct normalization factor. This yields $\beta=\frac{3}{4}m^{-3}$ and thus, an explicit continuous approximation of our p.m.f for large $n$ as

\begin{equation}
P_{\chi_n}(\alpha_i) = \sqrt{\frac{3}{\pi}}m^{-\frac{3}{2}}	\exp\left(-\frac{3\alpha_i^2}{4m^3}\right)
\label{Eq:Exact_Pchi}
\end{equation}

\noindent where $\alpha_i^2=(4i^2-4m^2i+m^4)$, $m=n/2$. A convenient form for finding the natural sufficient statistics for our distribution is given by dropping the index $i$, \textit{i.e.}  $\alpha_i = \alpha(\tau)$ so now $\chi$ is seen as a continuous variable. Allowing the mean back into the expression, we get 

\begin{equation}
f_{\chi}(\alpha;\mu_\chi,\sigma_\chi) = \frac{1}{\sigma_\chi\sqrt{2\pi}}	\exp\left(-\frac{(\alpha-\mu_\chi)^2}{2\sigma_\chi^2}\right),
\label{Eq:Asymptotic_gaussian_chi_pdf}
\end{equation} 

\noindent thus $f_{\chi}(\alpha;\mu_\chi,\sigma_\chi)$ is a probability density function that approximates $P_{\chi_n}(\alpha_i)$ for large $n$ and allows for the possibility of a change in location and scale. More importantly, Eq. (\ref{Eq:Asymptotic_gaussian_chi_pdf}) indicates directly that the natural sufficient statistics for estimating the mean and variance from a sample $\chi_1,\chi_2,\ldots,\chi_N$ are given by the sample mean and variance

\begin{eqnarray}
\bar{\alpha}&=&\frac{1}{N}\sum_{\tau\in T_L}\alpha(\tau) \\
s_\alpha^2&=&\frac{1}{N-1}\sum_{\tau \in T_L} \alpha(\tau)^2 \\ \nonumber
\label{Eq:SuffStats_alpha_s2}
\end{eqnarray}

\noindent where $T_L$ is the set of permutations corresponding to the ordinal patterns observed in a time series of length $L$ and $|T_L|= N = L - (n-1)l$ is the number of patterns observed in that realization. 
The sample mean $\bar{\alpha}$ is of special interest for statistical analysis, as we will show in the next section.





\begin{figure}[t!]
\centering
\includegraphics[width=0.51\textwidth]{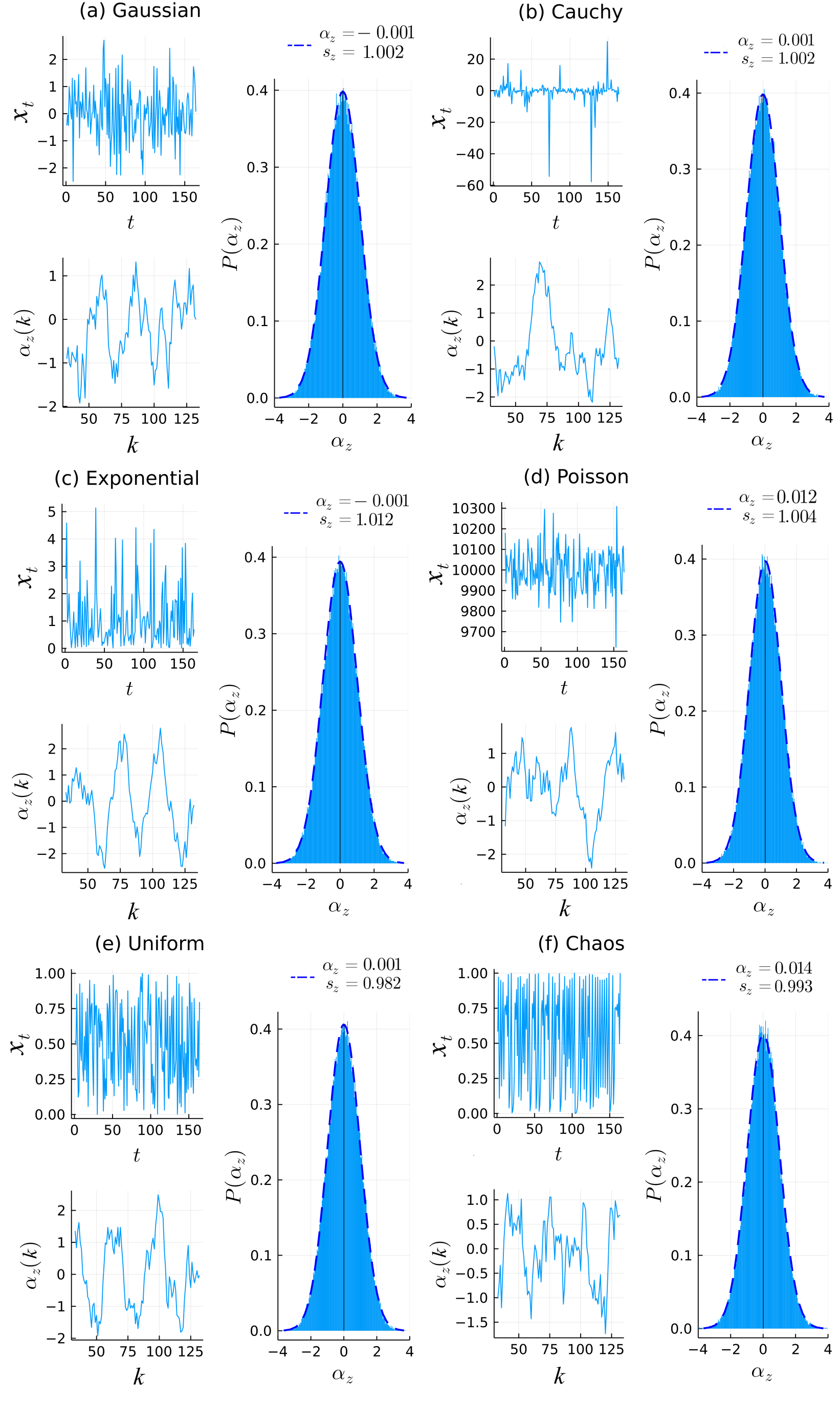}
\caption{White noise $\alpha$-analysis for white noise trajectories of length $L=10^5$, with source distributions (a) Standard Normal, (b) Cauchy($\mu=0$), (c) Exponential($\theta=1$), (d) Poisson($\lambda=10000$), (e) Continuous Uniform $U(0,1)$ and (f) a deterministic chaotic trajectory from the logistic map in Eq. (\ref{Eq:LMdef}) with $r=4$. The random trajectory is shown on the top left part of each panel, with the first 164 points displayed. Below these random realizations the first 132 points of the associated $\alpha$-process are shown. To the right of each sub panel we find the corresponding normalized histograms for $n=32$, $l=1$. Superimposed Gaussian densities with $\mu=\bar{\alpha}_z, \ \sigma^2 = s^2_{\alpha,z}$ are drawn with a blue dashed line. The sample mean $\bar{\alpha}_z$ is indicated by a black vertical line. } \label{Fig:alphaWNoises}
\end{figure}

\section{Time series analysis \label{Sec:TSA_StatInference} }

As an illustration, let us apply our framework as explained at the beginning of Sec. \ref{Sec:AlphaTau} to white noise time series of length $L = 10^5$ generated from different distributions: Standard Normal (unbounded support), Cauchy (unbounded support, heavy tails), Exponential (asymmetric distribution, unbounded support), Poisson (discrete unbounded support), Continuous Uniform (bounded support), and the special case of deterministic chaotic trajectories generated from the logistic map
\begin{equation}
f(x) = rx(1-x), \ \ x \in [0,1], \ \ r \in [0,4]
    \label{Eq:LMdef}
\end{equation}
at control parameter value $r=4$. At this value of $r$, the logistic map is ergodic and its invariant density has a closed form $\rho(x) = 1/\pi\sqrt{x(1-x)}$, which is equal to  $\mathrm{Beta}(1/2,1/2)$. Furthermore, its orbits display exponential decay of correlations \cite{Keller1992} and thus are effectively random in the long run. Therefore we can regard long time series obtained from \ref{Eq:LMdef} at $r=4$ as white noise generated from a Beta distribution as a source, \textit{i.e.}, $X \sim \mathrm{Beta}(1/2,1/2)$. As discussed in Sec. \ref{Sec:AlphaTau}, the validity of the theory developed for our functional requires the process $X_t$ to have a source distribution with continuous support, since a  total order is needed for obtaining well-defined ranking permutations, but this is not the case for the Poisson distribution. Nevertheless, here we relax this condition to see that, in a practical situation where the discrete support of $X_t$ is large enough to make the probability of repeated neighbouring points in time very low, then we can expect our method to apply to a good approximation.

For the sake of comparison, we use the standardized variable $Z_\alpha$, whose associated process is $\{ \alpha_{z,k} \}_{k\in [N]}$. Now let us choose an embedding dimension (sliding window length) with $n=32$ and let us use a lag of $l=1$. With this choice we ensure that the shape of $P_\chi(\alpha)$ is well approximated by a Gaussian (see Fig. \ref{Fig:pmf_n_4_8_16_32}-(d)).

In Fig \ref{Fig:alphaWNoises}, we can confirm empirically the statement that the only requirements for obtaining a Gaussian distribution for $\alpha(\tau)$, are the statistical independence in the observations and the continuity of the support of $X_t$. These conditions can be relaxed to include processes with sufficiently rapid (\textit{i.e.} exponential) decay of correlations as in the case of the deterministic chaotic trajectory, or to discrete processes with sufficiently large support.

\begin{figure}[t!]
\centering
\includegraphics[width=0.49\textwidth]{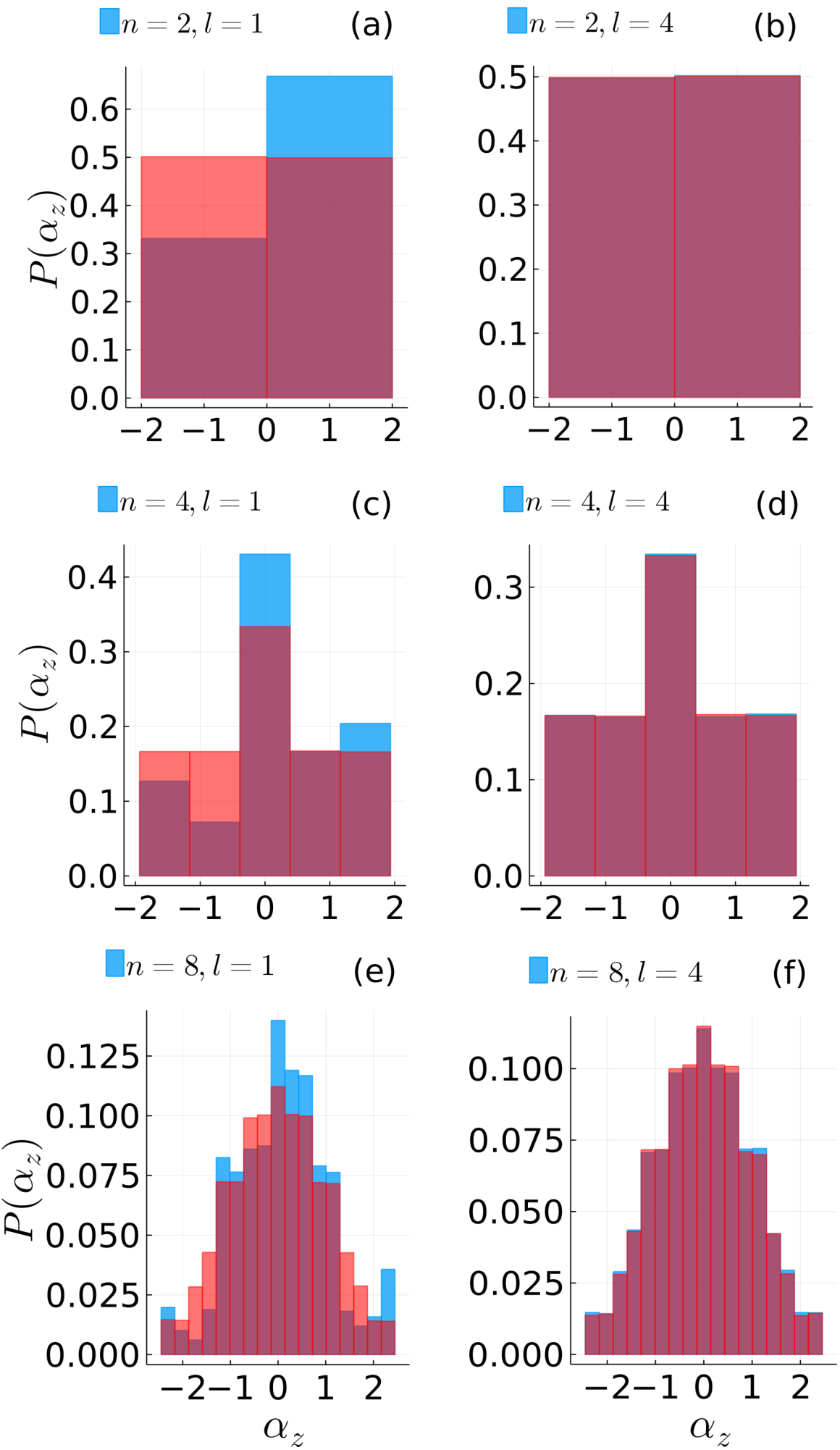}
\caption{Distributions $P(\alpha_z)$ for $n=2,4,8$ and different $l$ values, obtained for a chaotic trajectory of the logistic map with $r=4$ (blue bars), and white noise (red bars). The effect of the lag can be observed in panels (a),(c) and (e), where $l=1$ and the thus the dynamics of the map is reflected with more detail, including the effect of the forbidden patterns (see text). By contrast, when the lag is large enough, $l=4$ in this case, we see in panels (b),(d) and (f) how the effect of the correlations due to the determinism of the map, as well as effect of the forbidden patterns, is lost and both distributions are almost indistinguishable. 
} \label{Fig:Low_n_fingerprint_chaos}
\end{figure}

The usefulness of our analysis is not limited to large values of the embedding dimension. In Fig. \ref{Fig:Low_n_fingerprint_chaos} we show the distributions $P(Z_\alpha)$ for embedding dimensions $n=2,4,8$, obtained from a chaotic trajectory of length $L=10^4$ from the logistic map with $r=4$ and initial condition $x_0 = 0.84291157\ldots$ . The discrepancies between the distributions for the chaotic trajectory and realizations of uniform white noise come from the impossibility of the logistic map of displaying some types of patterns, known as forbidden patterns \cite{Amigo2007}. For instance, in \cite{Amigo2007} it is shown that the pattern of the type $\langle 3,2,1 \rangle$, and more generally patterns of the form $\langle *,3+k,*,2+k,*,1+k,* \rangle$ called outgrowth patterns (where $*$ indicates any other symbol in the pattern) cannot be displayed by orbits of the logistic map with $r=4$. Although one of the drawbacks of our method is that we are limited to even values of $n$, we can still see the effect of the forbidden patterns by the asymmetry in the distributions in Fig \ref{Fig:Low_n_fingerprint_chaos}, which has less mass in the negative side of the support. This is because the forbidden pattern has a negative value of our functional $\alpha(321) = -2$, and thus, patterns that are negative in the $\alpha$ sense will be more likely to belong to the outgrowth set patterns of $\langle 3,2,1 \rangle$. Correspondingly, an excess of positive patterns is observed, yielding an  asymmetric distribution. This makes our analysis potentially useful for detecting signatures of determinism in an observed process by direct comparison with white noise. As it can be seen in Fig. \ref{Fig:Low_n_fingerprint_chaos}, this effect is lost for a larger value of the lag due to the increased scale of observation and consequent loss of correlations.  

\subsection{3D Diffusion of Gold Nanoparticles \label{sSec:Au-NPs}}

\begin{figure}[t!]
\centering
\includegraphics[width=0.5\textwidth]{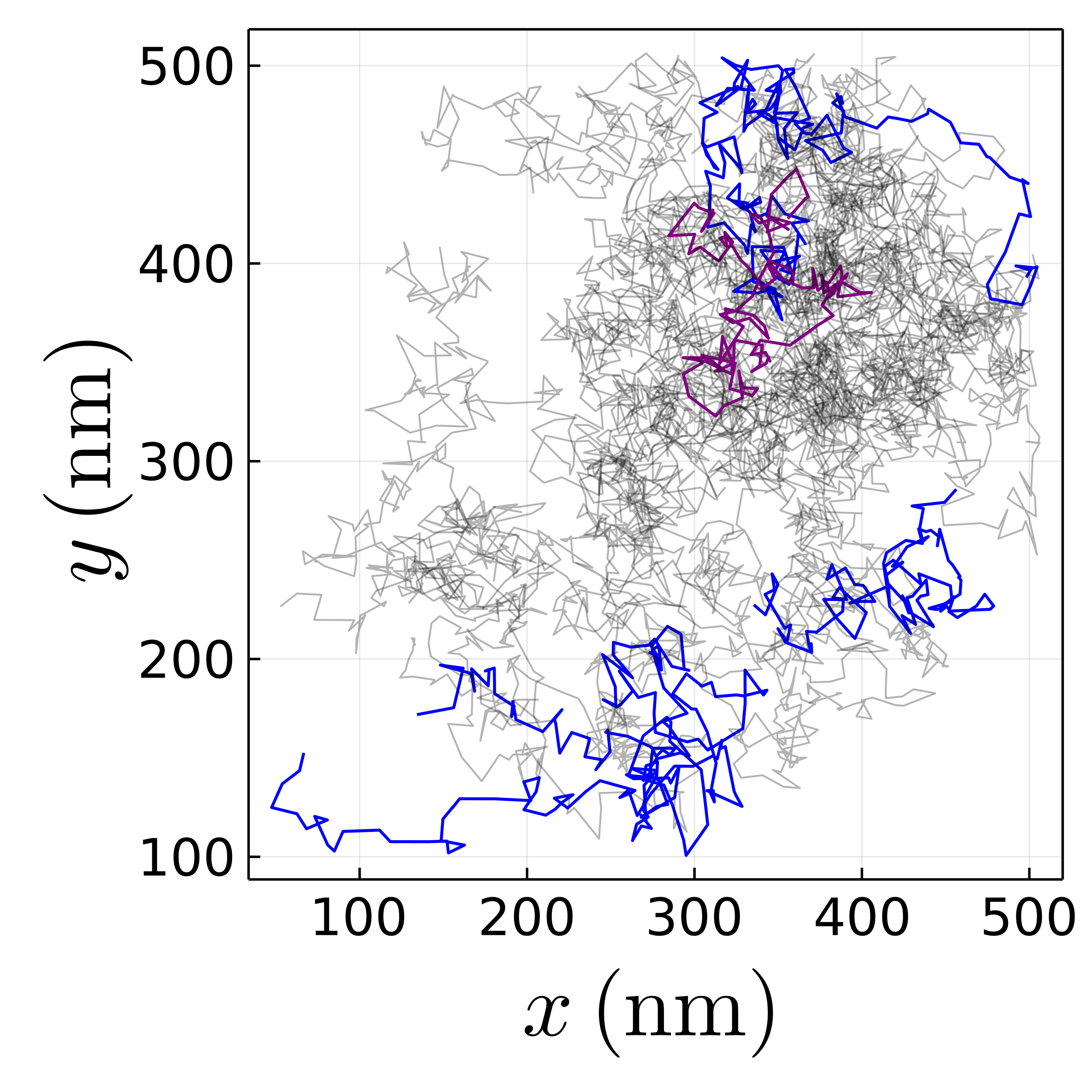}
\caption{Tracks of the 3D diffusion of $77$ nm gold nanoparticles in the $x$-$y$ plane. The 37 particles shown have a length $L>100$. The three tracks (labels 11,94 and 144) highlighted in blue did not pass the test of independence for the $x$-coordinate increments (see text). Analogously, the null hypothesis of the increments being  independent  in the $y$-coordinate had to be rejected for the highlighted purple track. The tracks for which the null hypothesis could not be rejected at the selected level of confidence are shown in gray.} \label{Fig:XYtracks}
\end{figure}

Now, let us analyze experimental data gathered using a recent and powerful technique for direct observation of the 3D dynamics of nanoparticles (NPs), known as liquid-cell scanning transmission electron microscopy (LCSTEM). Although it provides a very sharp resolution, this technique have been reported to yield observations of NP dynamics that is 3 to 8 orders of magnitude slower than the theoretical predictions \cite{Welling2020}. This discrepancy can be atributed to the damping effect of the strong beam of electrons, the viscosity of the media, and interactions of the particles with the boundaries of the experimental cell \cite{Welling2020}. In \cite{Welling2020}, Welling et \textit{al.} address the problem of observed slowed down diffusion by tuning the electron beam to a low dose rate and using high viscosity media, such as glycerol, for the NP diffusion. With those modifications, they track the 3D diffusion of charge-neutral 77 nm gold nanoparticles (Au-NPs) in glycerol as well as charged 350 nm titania particles in glycerol carbonate. The independence between the spatial increments is one of the defining properties of Brownian motion. In the following we show how to use our transformed ordinal pattern framework for testing this independence in the experimental particle tracks from the set of Au-NPs in Ref. \cite{Welling2020}. There are more than $200$ NP tracks observed in the $x$-$y$ plane in this data set, whose original experimental labels are kept here for identification. We analyzed all the trajectories whose length is $L \geq 100$ points (with a maximum of $L=359$), for a total of $M=37$ tracks  (See Fig. \ref{Fig:XYtracks}). 

The procedure is as follows. For each of the $M$ tracks $\mathbf{x} = (x_1,x_2,\ldots,x_L)$, $\mathbf{y}=(y_1,y_2,\ldots,y_L)$ 

\begin{itemize}
    \item[1)] Obtain the vectors of increments $\Delta\mathbf{x} = (x_2-x_1,x_3-x_2,\ldots,x_{L-1}-x_L)$ (analogously for $\Delta\mathbf{y}$). 
    
    \item[2)] Choose an embedding dimension $n$ and lag $l$ in order to apply the  $\alpha$-analysis to $\Delta\mathbf{x},\Delta\mathbf{y}$. This yields a pair of vectors, denoted for simplicity in notation by $\alpha\circ\Delta_j$ and $\alpha\circ\Delta_y$. The notation $\alpha\circ u $ has to be interpreted as first making an embedding of the series $u=\{u_1,u_2,\ldots u_L\}$ with the chosen $n,l$ values and then applying the $\alpha(\tau)$ functional to the corresponding collection of ranking permutations. 
     \item[3)] We get the standardized versions as 
     $$\alpha_z\circ\Delta_j=\frac{\alpha\circ\Delta_j}{\sigma_\chi} $$
    
    where $j=x,y$  and $\sigma_\chi = m((2m+1)/3)^{-1/2}$ (Eq. \ref{Eq:VarChi_exact_formula}). This amounts to obtaining the induced $\alpha_z$ time series for both coordinates, \textit{i.e.}, $\{\alpha_{z,k}^{(j)} \}_{k\in [N]}, \ \ j=x,y$.

    \item[4)] We now account for the correlations among the $\alpha_z$ values that are introduced by construction, by computing the effective length of the vectors  $\alpha_z\circ\Delta_j$ via \cite{Gelman2013} through the correction
     \begin{equation}
         N_{\mathrm{eff}} = \frac{N}{1+2\sum_j R_\alpha(i)}
         \label{Eq:Neff_lin_ACF_correction}
     \end{equation}
     where $R_\alpha(i)$ is the autocorrelation function (ACF)  of the $\{\alpha_{z,k}^{(j)} \}_{k\in [N]}$  series from the previous step, at time lag $i$.
    \item[5)] Finally, since the variance of $Z_\alpha$ is known ($\sigma_{Z_\alpha} = 1$), we can perform a one-sample $Z$-test on the $\{\alpha_{z,k}^{(j)} \}_{k\in [N]}, \ \ j=x,y$ series, under the assumption that the increments $\Delta\mathbf{x},\Delta\mathbf{y}$  are independent, and using $N_\mathrm{eff}$ computed in previous step as the sample size. 
    Therefore, our null hypothesis is simple: The mean value of the standardized variable $Z_\alpha$ is zero, $\mu_{Z_\alpha} = 0$. In other words, we want to test
    \begin{equation}
        H_0: \mu_{\alpha_z} = 0 \ \ \mathrm{against}  \ \ H_1: \mu_{\alpha_z} \neq 0
    \end{equation}
    with a selected level of confidence. 
\end{itemize}

The robustness of this test is guaranteed by the fact that the quantities $\bar{\alpha}$ and $s^2_\alpha$ (Eqs. \ref{Eq:SuffStats_alpha_s2}) obtained in Sec. \ref{Sec:SuffStats} are sufficient statistics for  $\chi$.  Therefore, we can rely on the statistic $\bar{\alpha}_z$ as the the unbiased estimator of the standardized mean $\mu_{Z_\alpha}$. For each spatial dimension of the diffusion, we have the estimators $\bar{\alpha}_{z}^{jw} = E[\alpha_z\circ\Delta_j], \ j=x,y$. We will follow common practice and choose a confidence level of $95\%$, corresponding to a type-I error (false-positive) rate of $0.05$, denoted here by $\epsilon_{1}$. This error rate is customarily denoted by ``$\alpha$'', conflicting with the notation for the main object in the paper. The author hopes that this change from the standard notation does not affect the reading. Correspondingly, we will denote the rate of a type-II error (false-negative) as $\epsilon_2$, customarily denoted by $\beta$, and will require an $80\%$ power, or $\epsilon_2=0.2$.
 
We choose an embedding dimension of $n=32$, since in that case we can consider $P(\alpha_z)$ to be well aproximated by a Gaussian (see Fig. \ref{Fig:pmf_n_4_8_16_32}-(d)). For the lag we choose $l=1$, because it ill give the maximum series length $N = L - (n-1)l$. The correlations introduced by this choice of lag are accounted for by Eq. (\ref{Eq:Neff_lin_ACF_correction}), but, before proceeding with next steps, let us estimate the minimum value of $N_{\mathrm{eff}}$ that complies with the chosen $\epsilon_1$ and  $\epsilon_{2}$. This estimate for $N_{\mathrm{eff}}$ is given from the usual two-sided $Z$-test by \cite{DeGroot2012} 
 
 \begin{equation}
      N_{\mathrm{eff}} \geq \left(z_{1-\frac{\epsilon_1}{2}} + z_{1-\epsilon_2} \right)^2\left(\frac{\sigma_{Z_\alpha}}{c}\right)^2,
      \label{Eq:Neff_min}
 \end{equation}

where $\sigma_{Z_\alpha}=\mathrm{Var}[Z_\alpha]$ (see Eq. (\ref{Eq:StandarsizedAlpha})), $z_{1-\epsilon_1/2}$ and $z_{1-\epsilon_2}$ are the quantiles of the standard normal distribution at  $1-\epsilon_1/2$ and $1-\epsilon_2$ respectively, and $c$ is the desired threshold of detection for deviations from the mean, regularly written as proportional to the standard deviation, so $c \sim \sigma$. 

For $\epsilon_1 = 0.05$, $\epsilon_2=0.2$, and $c = 1.96\sigma$,  we get  $N_{\mathrm{eff}}\geq 30$ from Eq. (\ref{Eq:Neff_min}). All of the trajectories considered have a corresponding $N_{\mathrm{eff}}>30$, therefore we can reliably apply our test. A summary of the results of the test is provided in Table \ref{Tab:SingleTrack_Ztest}, where we show the tracks for which the null hypothesis is rejected ($p$-value lower than $0.05$). 


\begin{figure}[t!]
\centering
\includegraphics[width=0.35\textwidth]{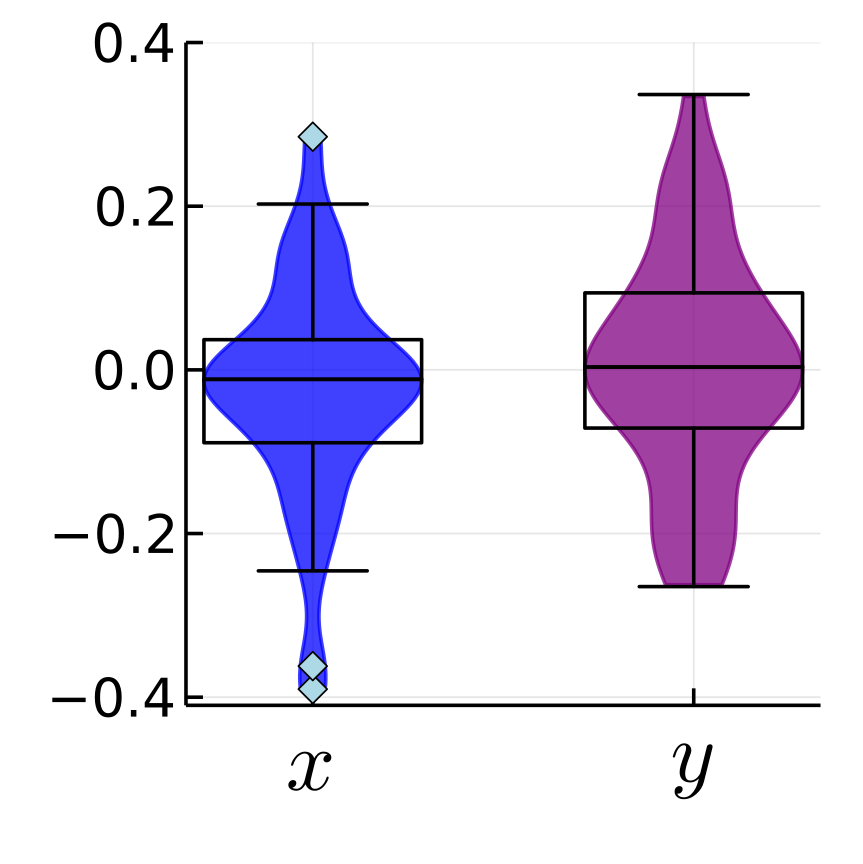}
\caption{Violin-box plot for the sample means $ \{\bar{\alpha}_z^{(j)} \}, \ j=x,y$ of the 37 analyzed tracks. The whiskers mark the 1.5$\times$IQR (inter-quartile range). Outliers are detected only in the $x$-coordinate $\bar{\alpha}_z$ set, corresponding to the trajectories with labels $11,94,144$ and indicated by diamonds.} \label{Fig:Violin-Box_plot}
\end{figure}

 In contrast with the analysis displayed in Fig. \ref{Fig:Violin-Box_plot}, where only 3 $x$-trajectories are detected as outliers (labels $11,94,144$), there are $4$ trajectories that get rejected by the single trajectory hypothesis test (See Table \ref{Tab:SingleTrack_Ztest}), with labels $11,94,144$ ($x$-tracks) and $159$ ($y$-track).

Nevertheless, the agreement between the single trajectory $Z$-tests and the independent analysis through the box plot is reasonably good, suggesting that the correction in Eq. (\ref{Eq:Neff_lin_ACF_correction}) represents a sensible approximation that accounts for the autocorrelation in the $\alpha_{z,k}^{(j)}$ processes.


\begin{table}[ht!]
\begin{center}
 \begin{tabular}{|c|c|c|c|c|c|c|}
 \hline
  NP label  & $x$, $p$-val & $y$, $p$-val & $\bar{\alpha}_{z}^{(x)}$ & $\bar{\alpha}_{z}^{(y)}$ & $N_{\mathrm{eff},x}$ & $N_{\mathrm{eff},y}$\\ 
 \hline\hline
 11  & 0.000 & 0.963 & -0.3902 & 0.0051 & 81 & 81\\ 
 \hline
 94  & 0.024 & 0.089 & 0.2849 & -0.2128 & 63 & 64\\ 
 \hline
 144  & 0.032 & 0.667 & -0.3620 & 0.0717 & 35 & 36\\ 
 \hline
 159  & 0.276 & 0.031 & -0.1700 & 0.3364 & 41 & 41\\ 
 \hline

\end{tabular}
\end{center}
\caption{Nanoparticle tracks rejected by the single trajectory analysis and their $\alpha$-statistics.}
\label{Tab:SingleTrack_Ztest}
\end{table}







\section{Discussion \label{Sec:Discussion}}

We have illustrated the main advantages of avoiding working with the direct statistics of patterns by, instead, first dividing the symmetric group into classes through the functional $\alpha(\tau)$ defined by Eq. (\ref{Eq:AlphaTau_Def_even_n}), so that the problem is reduced to analyze these classes. However, this procedure does not come without drawbacks, and next we will discuss this, as well as other positive points in more detail. 

An interesting conceptual consequence of the presented view of white noise is, that a sense of typicality emerges in terms of the functional $\alpha(\tau)$ due to its concentration around zero. Therefore, the stationarity of white noise acquires a combinatorial character arising from statistical constraints.

In Sec. \ref{Sec:TSA_StatInference}, we have successfully shown a use case of our framework to test for independence in the spatial increments of diffusing particles in 3 dimensions, whose motion was recorded in a 2-dimensional plane. Although computing the ensemble averaged mean squared displacement (MSD) is the customary check for diffusive behavior, it does not provide the single-trajectory detail and statistical power of our method. Indeed, our method can be implemented for a single particle if that is the only information available, and still being reliable assuming a minimum effective trajectory length is achieved (as seen in Sec. \ref{sSec:Au-NPs}), which is a sensible requisite.

The time series available were of a relatively short length. Yet, notably, the test is able to handle these short time series. None of the trajectories were rejected by our test for the two dimensions at the same time. That could be interpreted as a good indication that the observation technique used in \cite{Welling2020} performed well enough as to preserve the 3D Brownian diffusion overall, by keeping the introduction of correlations in the motion through the observation scheme at a minimum. After discussion with one of the authors in \cite{Welling2020}, we can explain the higher rejection rate for the $x$-coordinate tracks by the observation procedure: The LCSTEM probe is progressively scanning line by line along the $x$-dimension, thereby potentially introducing weak correlations in the particles motion in that direction.

The former is supported from the theory exposed and the reasonably good agreement between our analysis and the quartile analysis in Fig. \ref{Fig:Violin-Box_plot}. Furthermore, the quartile analysis did not determined a track whose $y$-component was rejected by our test, suggesting our method is more sensitive for a given confidence level. 

For all the examples of white noise processes considered in Sec. \ref{Sec:TSA_StatInference}, the customary ordinal pattern analysis for the PE computation would be practically impossible for the choice of $n=32$ used here, since we have to keep track of $32!$ patterns. The analysis and graphical representation of the final distribution would be also impossible without a further coarsening of the support, something that would render the analysis crippled of the detail that characterizes it in the first place. Instead, in the present framework the size of the support of our distribution is $|A_n|=m^2+1$ (see Sec.\ref{Sec:OrdPatterns}). This is an important simplification, while still keeping relevant information about the patterns both in terms of correlations, and even rough information about the variation of the amplitudes in the form of weights. A remarkable aspect of our framework when comparing white noise of different sources is the robustness of the empirical distributions. The empirical distribution over $S_n$ in the customary PE approach approximates a discrete uniform with support $\{1,2,\ldots,n!\}$, implying that for moderate to large $n$, it would display strong variations when estimated from a finite sample. In the considered case with $n=32, \ l=1$ and series length $L=10^5$, the number of observations $N = L-(n-1)l \simeq 10^5$ falls extremely short for having at least one representative out of the $32! \simeq 2.6\times10^{35}$ possible patterns. Therefore the obtained empirical density would be composed mostly of void regions and uneven peaks. In order to prevent this effect, in our current example of time series of length $L=10^5$ one should limit to $n=8$ ($8! \simeq 4\times10^4$) and still, we could get an uneven empirical density with gaps despite the rather large time series. In contrast to this, in our approach there are just $m^2+1=257$ classes to keep track of, as in the illustrative examples considered in Fig. \ref{Fig:alphaWNoises}. The choice $n=32$ was done mainly to guarantee a good approximation of $P(Z_\alpha)$ by a Gaussian p.d.f. Nevertheless, we can think of alternative analysis for lower $n$ values exploiting the fact that we know the specific form of $P(Z_\alpha)$ for hypothesis testing. 


Now, let us address the autocorrelations introduced in the embedding procedure. We already mentioned in Sec. \ref{Sec:AlphaTau} that the sequence of $\alpha$ values obtained from a time series are correlated due to the overlap among the embedding vectors, that in turn translates into overlapping ranking permutations $\tau\in S_n$ and, finally, correlated $\alpha$ values. This is specially so for low values of the lag $l$. As illustrated in Sec. \ref{Sec:TSA_StatInference}, for large enough values of $l$, the effect of the correlations in the original process can be significantly diminished, but also the overlap between the patterns can be diminished or eliminated (see Fig.\ref{Fig:Low_n_fingerprint_chaos}). Nevertheless, even for low $l$, the autocorrelations in the series $\{\alpha_k\}_{k\in [N]}$ do not affect the Gaussian character of $P(\alpha)$, since this characteristic comes from the fact that, as $n\rightarrow\infty$, the partial sums $s_l$ discussed in Sec. \ref{Sec:AlphaTau} are composed of integers from the uniform distribution, that become effectively independently sampled as $n$ grows, and thus the Central Limit Theorem applies. This is the same effect as the effective loss of statistical dependence when drawing with replacement from a very large pool. Thus, despite the correlations introduced by construction in the method, this does not come at a cost so high that it ruins the statistical power of the analysis, specially for large $n$, large $l$, or both $n,l$ large values.



The absence or over representation of patterns that is expected to happen due to finite sampling is lessened by the fact that it is very likely that a pattern in the same or a similar class will take the place of the missing one. Or vice-versa,  over represented patterns in a sample would induce over representation of patterns in neighboring classes for low lag values, specially $l=1$ when the window has the greatest overlap. This has the overall effect of perturbing the shape of the Gaussian around the over represented pattern. It is a similar situation for absent patterns. An example of this can be seen for the chaotic trajectory of the logistic map in Fig. \ref{Fig:alphaWNoises}-(f), where there is a region in the center which gets a slightly increased probability density than it should for a white noise process. This is explained by the missing forbidden patterns as discussed in Sec. \ref{Sec:TSA_StatInference}. The over representation of the patterns with values of $\alpha$ around zero is apparently more likely to happen for the processes with bounded support as is the case of Uniform white noise and the deterministic chaotic trajectory (See Fig \ref{Fig:alphaWNoises}(e)-(f)). This could be explained by the boundeness of the noise and the finiteness of the trajectory, making less likely that increasing (decreasing) sequences appear, corresponding to  positive (negative) values of $\alpha$ that are far from the average around zero. Thus, values $\alpha \simeq 0$ get over represented.



A major drawback for the applicability of our framework is, that an even embedding dimension must be used. Nevertheless, a workaround to this could be to perform the analysis for the adjacent even values of the desired odd $n$ value if the actual odd structure of the patterns is not relevant. Furthermore, in principle the odd $n$ analysis is also possible by the Definition 1 (Eq. (\ref{Eq:AlphaTau_Def})) that we can approximate the exact distribution of the corresponding $\alpha$ functional since we have the general expression \ref{Eq:Def_GaussianBinomial_Bona} describing the statistics of the degeneracy of $\alpha$ for any choice of $n$ and $m$. This degeneracies can be computed numerically from the expression \ref{Eq:Def_GaussianBinomial_Bona} by means of a recursion for the Gaussian binomial coefficients \cite{CombPerm_Bona2012}. The distributions for $\alpha(\tau)$ for odd $n$ thus obtained are skewed, but still bell-shaped. We wish to make a general analysis of this case, together with possible practical implications in a future work. 


To finish, we stress that an important message conveyed with this contribution is, that the full detail of the customary ordinal pattern analysis (prior to PE computation) is not needed and instead hinders its statistical applications and, on the other hand, that the computation of the Shannon entropy directly from the ordinal pattern analysis washes away information that is valuable statistically. The approach presented here represents a middle ground with several extra benefits and relatively minor drawbacks. 

\begin{acknowledgments}
Funding for this work was provided by the Alexander von Humboldt Foundation in the framework of the Sofja Kovalevskaja Award endowed by the German Federal Ministry of Education and Research, through Matteo Smerlak. The author acknowledges Tom Welling,  Marijn A. van Huis, Professor van Blaaderen and their collaborators for sharing their experimental data. David Davalos, Holger Kantz and Mario Díaz-Torres are acknowledged for useful discussions and comments. 
\end{acknowledgments}

\bibliography{AlphaTau_Refs}

\appendix

\section{Continuous approximation for finite permutation length\label{App:A}}

It is clear from Fig. \ref{Fig:pmf_n_4_8_16_32}, that the p.m.f $P(\alpha_z)$ is not well approximated by a Gaussian for low values of $n$. Therefore, we propose a Beta distribution as a better continuous approximation  for this case. Recall its probability density function (p.d.f.) 

\begin{equation}
f_W(w;a,b) = \frac{w^{a-1}(1-w)^{b-1}}{B(a,b)}, \ \  w \in [0,1]
\label{Eq:Beta_pdf}
\end{equation}

\noindent where $B(a,b) = \int_{0}^{1} t^{a-1}(1-t)^{b-1}dt$.

To match the random variable $\chi$ with the Beta random variable $W$, we need first to transform the support $A_n$ of $P_\chi(\alpha)$ in Eq. (\ref{Eq:alpha-support}) to be a subset of the unit interval $\tilde{A}_n \subseteq [0,1]$ by a shift and re-scaling 

\begin{equation}
W_\alpha = \frac{\chi+m^2}{2 m^2}, \  \ W_\alpha \sim f(\alpha_w;a,b)
\label{Eq:W_alpha_def}
\end{equation}

 where $f(\alpha_w;a,b)$ is a Beta p.d.f. (Eq. \ref{Eq:Beta_pdf}). As in the case of $Z_\alpha$, here we adopt the notation $W_\alpha$ for the transformed random variable, $\alpha_w$ for its realizations and $P(\alpha_w)$ for the p.m.f..

\noindent From Eq. (\ref{Eq:W_alpha_def}), formula (\ref{Eq:VarChi_exact_formula}) and $\mu_\chi=E[\chi]=0$, we get for the moments of $W_\alpha$

\begin{eqnarray}
E[W_\alpha] = \mu_{W_\alpha} &=&\frac{1}{2} \\
\mathrm{Var}[W_\alpha] = \sigma^2_{W_\alpha} &=& \frac{\sigma^2_{\chi}}{4m^4} = \frac{2m+1}{12m^2},
\label{Eq:W_alpha_moments}
\end{eqnarray}

This low $n$ approximation of $P_{W_\alpha}(\alpha_w)$ by a Beta distribution $f(\alpha_w;a,b)$ is consistent with the fact that in the limit $a,b \rightarrow \infty$ the Beta p.d.f. in Eq. (\ref{Eq:Beta_pdf}) converges to a Gaussian, as seen in Sec \ref{Sec:SuffStats}.

\noindent The random variable $W_\alpha \in [0,1]$ can be interpreted as a probability itself. A value $\alpha_w(\tau)$ would correspond to the probability of a pattern $d_\tau$ to have a degree of asymmetry ranging from $0$ to $1$. This means that $\mathrm{Pr}(0 \leq \alpha_w < \frac{1}{2})$ is the probability of sampling an overall decreasing pattern, $\mathrm{Pr}(\alpha_w = \frac{1}{2})$ is the probability of sampling an overall constant pattern, and $\mathrm{Pr}( 1 \geq \alpha_w > \frac{1}{2})$ is the probability of finding overall increasing patterns. 



\noindent The Chebyshev-Pearson method of moments \cite{ShaoBook} provides the following expressions for estimating the parameters of the $B(a,b)$  distribution from the mean $\bar{x}_w = N^{-1} \sum_j^N \alpha_{w,j}$ and variance $\hat{\sigma}_w^2 = (N-1)^{-1}\sum_j^N(\alpha_{w,j}-1/2)$ from a random sample  $W_{\alpha,1},W_{\alpha,2}\ldots W_{\alpha,N}$

\begin{eqnarray}
\hat{a} &=& \bar{x}_w\left(\frac{\bar{x}_w(1-\bar{x}_w)}{\hat{\sigma}_w^2} -1 \right) \\
\hat{b} &=& (1-\bar{x}_w)\left(\frac{\bar{x}_w(1-\bar{x}_w)}{\hat{\sigma}_w^2} -1 \right) \\ \nonumber
\label{Eq:MethodMoments_BetaParams}
\end{eqnarray}

\noindent Since in our computations the sample comprises the population itself (the whole symmetric group) the parameters we have characterized so far are the exact population mean and the exact variance, respectively, \textit{i.e.,} $\bar{x}=\mu_{W_\alpha}=1/2$, $\hat{\sigma}=\sigma_{W_\alpha} $, $a =\hat{a}$, $b = \hat{b}$.  Then, replacing Eqs. (\ref{Eq:W_alpha_moments})  in formulas (\ref{Eq:MethodMoments_BetaParams}) to obtain the exact parameters 

\begin{equation}
 a =  \frac{3m^2}{2(2m+1)} - \frac{1}{2} , \ \mathrm{and} \ \ b = a
 \label{Eq:BetaParams}
\end{equation}

\noindent that becomes $a\sim \frac{3}{4}m$ for $m\gg 1$, in agreement with the symmetry of $P(\alpha)$ and the  linear increase observed for the parameters $a, b$ from the numerical computations. For the case $m=1$ we get $a=b=0$. In this limit, the p.d.f. for the Beta distribution in Eq. (\ref{Eq:Beta_pdf}) becomes a Bernoulli distribution with $p=1/2$, which is precisely the same as  $P_{W_\alpha}(\alpha_w)$ for $n=2$, or, after a location shift to zero, is also equivalent to  $P_n(\alpha)$ and $P(\alpha_z)$ which are symmetric Bernoulli distributions.  

\section{Concentration of measure by $\alpha(\tau)$ \label{App:B}}

Before drawing a conclusion on the continuous asymptotic shape of  $P_\chi(\alpha)$, let us characterize its moments.

\noindent First, we find a bound for the scaling of $\sigma_\chi$ in order to test  formula \ref{Eq:VarChi_exact_formula} through the following sub-gaussianity argument. 

A random variable $X$ with finite expectation $\mu$ is called sub-gaussian  if there exists a positive number $\sigma$ such that  

\begin{equation}
E[\exp(\lambda(X-\mu))] \leq \exp \left( \frac{\lambda^2\sigma^2}{2} \right), \ \forall \lambda \in \mathbb{R}.
\label{Eq:Def_sub-gaussian1}
\end{equation}

\noindent The constant $\sigma$ is called proxy variance or sub-gaussian norm \cite{Marchal2017}. From the classical Hoeffding's theorem \cite{Hoeffding1963} together with the definition in Eq. (\ref{Eq:Def_sub-gaussian1}), we can conclude that every integrable random variable supported on a compact set is sub-gaussian \cite{Marchal2017}. Therefore, considering that for fixed $n$ our functional is bounded $-m^2 \leq \alpha \leq m^2$, we conclude that $\chi$ is subgaussian, and so it is $Y$. \\

\noindent The left-hand side of inequality (\ref{Eq:Def_sub-gaussian1}) is nothing more than the moment generating function $M_{X'}(\lambda) = E[\exp(\lambda X')]$  of the random variable $X' = X-\mu$. If we apply of Markov's inequality to $M_{X'}(\lambda)$ we obtain  

\begin{eqnarray}
\mathrm{Pr}(X'\geq t)&=& P(\exp(\lambda X') \geq \exp(t\lambda)) \nonumber \\
           &\leq &  \frac{E[\exp(tX')]}{\exp(t\lambda)}
\label{Eq:MarkovMGF} 
\end{eqnarray}

by finding the minumum in the right-hand side of the inequality in Eq. (\ref{Eq:MarkovMGF}) with respect to the parameter $\lambda$ we get the optimal bound (Chernoff bound \cite{DeGroot2012},\cite{Chernoff1952}) for subgaussian random variables \cite{Marchal2017}

\begin{equation}
\mathrm{Pr}(X-\mu> t)\leq \exp \left(-\frac{t^2}{2\sigma^2} \right), \ \forall t>0,
\label{Eq:Def_sub-gaussian2}
\end{equation} 

\noindent which is an equivalent but more usable statement of the subgaussianity property, Eq. (\ref{Eq:Def_sub-gaussian1}). Although inequalities (\ref{Eq:MarkovMGF}) and (\ref{Eq:Def_sub-gaussian2}) are more useful when treating sums of i.i.d variables due to factorizability, they are valid also for certain instances of dependency in the samples, such as sampling without replacement \cite{Hoeffding1963}. Computing $\alpha(\tau)$ is a problem of sampling $m=n/2$ symbols without replacement from the set $[n]$, so we can apply these techniques in our problem. 
Considering that $\mu_X=0$ and the fact that we always know the tail probability of $P(\chi)$ 

\begin{eqnarray}
\mathrm{Pr}(|\chi| \geq m^2) &=& p_0 + p_1 + p_{m^2-1} + p_{m^2} \nonumber \\
			 &=& \frac{4}{{{n}\choose{m}}}  
\label{Eq:Tail_prob}
\end{eqnarray}

\noindent as stated in the previous section.  Using Eq. (\ref{Eq:Def_sub-gaussian2}) with $t=m^2$ together with Eq. (\ref{Eq:Tail_prob}), taking logarithm in both sides and using Strling's approximation for the logarithm of the central binomial coefficients up to the first term: $\log{{n}\choose{m}}\simeq m\log(4)$, and recalling $n=2m$, we get the bound

\begin{equation}
\sigma_\chi^2 \leq \frac{m^3}{\left[\left(1+\frac{1}{m}\right)\log(4)-\frac{1}{2m}\log(\pi m) \right]}
\label{Eq:VarXfullBound}
\end{equation}

 which reduces rapidly to

\begin{equation}
\sigma_\chi^2 \leq \frac{m^{3}}{\log(4)}
\label{Eq:VarXbound}
\end{equation}

\noindent as $m\rightarrow\infty$, which is consistent with formula  (\ref{Eq:VarChi_asymptotic}) for $m \geq 7$.  \\

\end{document}